# A Life Expectancy Study based on the Deterioration Function and an Application to Halley's Breslau Data

## Christos H Skiadas

Technical University of Crete, Data analysis and forecasting laboratory, Chania, Crete, Greece.
E-mail: skiadas@cmsim.net

**Abstract:** Further to the proposal and application of a stochastic methodology and the resulting first exit time distribution function to life table data we introduce a theoretical framework for the estimation of the maximum deterioration age and to explore on how "vitality," according to Halley and Strehler and Mildvan, changes during the human lifetime. The mortality deceleration or mortality leveling-off is also explored. The effect of the deterioration over time is estimated as the expectation that an individual will survive from the deterioration caused in his organism by the deterioration mechanism. A method is proposed and the appropriate software was developed for the estimation of life expectancy. Several applications follow.
The method was applied to the Halley life table data of Breslau. Extrapolations are done showing a gradual improvement of vitality mechanisms during last centuries.

## Introduction

The methods and techniques related to life expectancy started to grow from the proposal of Life Tables' system by Edmond Halley (1693) in his study on Breslau birth and death data. The related methods include mathematical models starting from the famous Gompertz (1825) model. The Gompertzian influence on using the mortality data to construct life tables was more strong than expected. He had proposed a model and a method to cope with the data. The model was relatively simple but quite effective because he could, by applying his model, to account for the main part of the data set. The method he proposed and applied was based on a data transformation by adding, dividing and finally taking logarithms of the raw data in an approach leading to a linearization of the original distribution. This was very important during Gompertz days when the calculations where very laborious. Furthermore, this method of data linearization made possible the wide dissemination, approval and use by the actuarial people of the Gompertz model and the later proposed Gompertz–Makeham (Makeham, 1860) variation. However, the use of the logarithmic transformation of the data points until nowadays turned to be a strong drawback in the improvement of the related fields. By taking logarithms of a transformed form of the data it turns to have very high absolute values for the data points at the beginning and at the end of the time interval, the later resulting in serious errors due to the appearance of very high values at high ages. To overcome these problems resulting from the use of logarithmic transformations several methods and techniques have being proposed and applied to the data thus making more complicated the use of the transformed data. The approaches to find the best model to fit to the transformed data leaded to more and more complicated models with the Heligman-Pollard (1980) 8-Component Model and similar models to be in use today. The task was mainly to find models to fit to data well, instead to search for models with a good explanatory ability. Technically the used methodology is directed to the actuarial science and practice and to applications in finance and insurance (see the method proposed by Lee and Carter, 1992).

Another quite annoying thing by applying logarithms is that the region around the maximum point of the raw data death distribution including this inflection point along with the left and right inflection points are all on the almost straight line part of the logarithmic curve.

From the modelling point of view it is quite difficult to find from the transformed data the characteristics of the original data distribution. Even more when constructing the life tables the techniques developed require the calculation of the probabilities and then the construction of a "model population" usually of 100000 people. The reconstruction of the original data distribution from the "model population" provides a distribution which is not so-close to the original one.

Here we propose a method to use models, methods and techniques on analysing data without transforming the data by taking logarithms. While this methodology has obvious advantages it faces the problem of introducing it in a huge worldwide system based on the traditional use of methods, models and techniques arising from the Gompertzian legacy. To cope with the well established life table data analyses we present our work as to test the existing results, to give tools for simpler applications and more important to make reliable predictions and forecasts. We give much attention to the estimation and analysis of the life expectancy and the life expectancy at birth as are the most important indicators for policy makers, practitioners and of course researchers from various scientific fields.

**The Deterioration Function: Further analysis**

The Deterioration Function for the model $H(t) = l - (bt)^c$ or for the model $H(t) = (bt)^c$ is expressed by the following formula (Skiadas, 2011). This formula provides the value of the curvature at every point $(H(t), t)$.

$$K(t) = \frac{|c(c-1)b^c t^{c-2}|}{(1 + c^2 b^{2c} t^{2c-2})^{3/2}}$$

This is a bell-shaped distribution presented in Figure 1. The first exit time IM-model (Skiadas, 2007, 2010a, 2010b, 2011 and Janssen and Skiadas, 1995) is applied to the female mortality data of Italy for the year 1950.

$$g(t) = \frac{k(l + (c-1)(bt)^c)}{\sqrt{t^3}} e^{-\frac{(l-(bt)^c)^2}{2t}}$$

A simpler 3 parameter version of this model arises when the infant mortality is limited thus turning the parameter $l$ to be: $l=0$ and the last formula takes the simpler form:

$$g(t) = \frac{k(c-1)(bt)^c}{\sqrt{t^3}} e^{-\frac{(bt)^{2c}}{2t}}$$

The data, the fitting and the deterioration curves are illustrated in Figure 1. The deterioration function starts from very low values at the first stages of the lifetime and is growing until a high level and then gradually decreases. As the main human characteristics remain relatively unchanged during last centuries it is expected that the deterioration function and especially the maximum point should remain relatively stable in previous time periods except of course of the last decades when the changes of the way of living and the progress of biology and medicine tend to shift the maximum deterioration point to the older age periods.

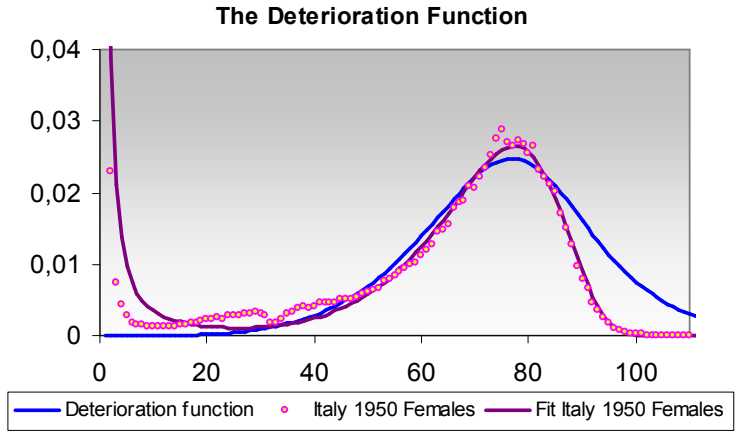

Fig. 1. Deterioration function, raw and estimated data for Italy.

We can also explore the Greenwood and Irwin (1939) argument for a late-life mortality deceleration or the appearance of mortality plateaus at higher ages by observing the shape of the deterioration function. As we can see in Figures 1 and 2 the deterioration tends to decrease in higher ages and especially after reaching the maximum value, leading to asymptotically low levels thus explaining what we call as Mortality Leveling-off. The deterioration of the organism tends to zero at higher ages. Economos (1979, 1980) observed a mortality leveling-off in animals and manufactured items.

The characteristic non-symmetric bell-shaped form of the deterioration function is illustrated in Figure 2 for females in France and for various periods from 1820 to 2007. The IM-model is applied to the data provided by the human mortality database for France in groups of 10 years. The form of the deterioration function is getting sharper as we approach resent years. The improvement of the way of living is reflected in the left part of the deterioration function. The graph for 1820-1829 presents the deterioration starting from very early ages. Instead the graph for the period 1900-1909 shows an improvement in the early deterioration period. The improvement continues for 1950-1959 and 2000-2007 periods.

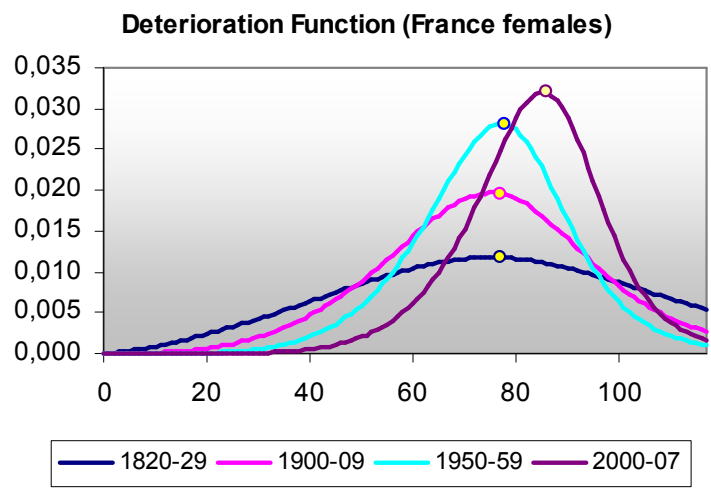

Fig. 2. Deterioration function for females in France

However, the maximum point of these curves is achieved at almost the same year of age from 1820-1829 until the 1950-1959 period. This is an unexpected result as it was supposed that the general improvement of the way of living would result in a delay in the deterioration mechanisms. To clarify this observation we will further analyze the deterioration function.

A main characteristic of the deterioration function is its maximum point achieved at:

$$T_{Deter} = \left[ \frac{(c-2)}{(2c-1)c^2 b^{2c}} \right]^{\frac{1}{2c-2}}$$

We will apply this formula to mortality data of several countries and for various time periods. The task is to explore our argument for the stability of the maximum deterioration point around certain age limits and of a shift of this point to higher ages.

Figure 3 for females in France from 1816 to 2007 is quite interesting. Four graphs are illustrated. The data are summarized in Table I.

**TABLE I**

Life expectancy at birth and Maximum deterioration age

France, females

| Year | HM-database | IM-model | 3p-model | Max Det |
|------|-------------|----------|----------|---------|
| 1816-1819 | 39,54 | 34,69 | | 74,79 |
| 1820-1829 | 39,58 | 34,98 | | 75,20 |
| 1830-1839 | 39,66 | 37,39 | | 75,36 |
| 1840-1849 | 41,19 | 39,15 | | 75,67 |
| 1850-1859 | 40,43 | 39,67 | | 75,43 |
| 1860-1869 | 42,27 | 39,74 | | 75,11 |
| 1870-1879 | 42,33 | 41,51 | | 75,63 |
| 1880-1889 | 44,81 | 42,54 | | 75,45 |
| 1890-1899 | 46,72 | 45,77 | | 74,98 |
| 1900-1909 | 49,83 | 48,87 | | 75,23 |
| 1910-1919 | 51,53 | 54,06 | | 75,53 |
| 1920-1929 | 56,41 | 55,41 | | 75,79 |
| 1930-1939 | 60,89 | 60,13 | | 76,06 |
| 1940-1949 | 62,00 | 61,50 | | 76,37 |
| 1950-1959 | 71,16 | 66,68 | 73,72 | 77,15 |
| 1960-1969 | 74,59 | 71,60 | 75,78 | 78,61 |
| 1970-1979 | 76,89 | 75,25 | 77,54 | 79,86 |
| 1980-1989 | 79,45 | 78,75 | 80,04 | 81,24 |
| 1990-1999 | 81,81 | 81,96 | 82,70 | 83,51 |
| 2000-2007 | 83,47 | 82,87 | 83,27 | 85,26 |

The data for the life expectancy at birth (blue line) are collected from the human mortality database for France (females in 10 year groups). The life expectancy at birth estimates based only on the fitting estimations of the IM model (based on the death data only) is presented with a magenta line. As the infant mortality becomes negligible during last decades we can estimate the life expectancy at birth by based on a 3-parameters alternative of the IM-model (red line). Note that according to the Nathan Keyfitz (2005) theory for a stable population the life expectancy estimates tend to coincide when using different methods. Here the three different estimates for the life expectancy at birth are similar for the last 20 years. That is interesting is that the life expectancy at birth is approaching the maximum deterioration age (light blue line) which may be seeing as a plateau for the life expectancy at birth.

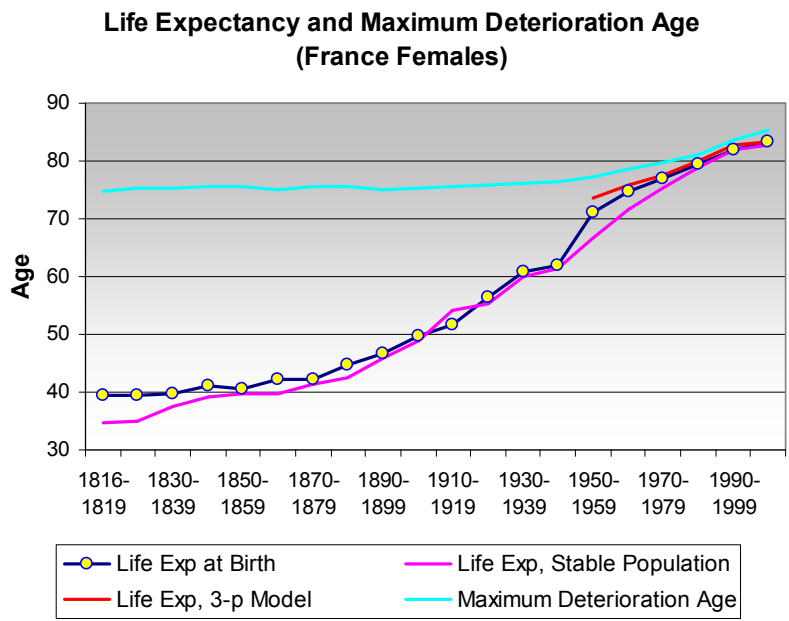

Fig. 3. Life expectancy at birth and maximum deterioration age for France.

Figure 4 illustrates the life expectancy at birth (blue line) with data collected from the human mortality database (females in 10 year groups). As the infant mortality becomes negligible during last decades we can estimate the life expectancy at birth by based on a 3-parameters alternative of the IM-model (red line). The maximum deterioration age is expressed by a cyan line. The results for the four countries studied, A) Netherlands, B) Denmark, C) Italy and D) Norway are similar to the previous application for France.

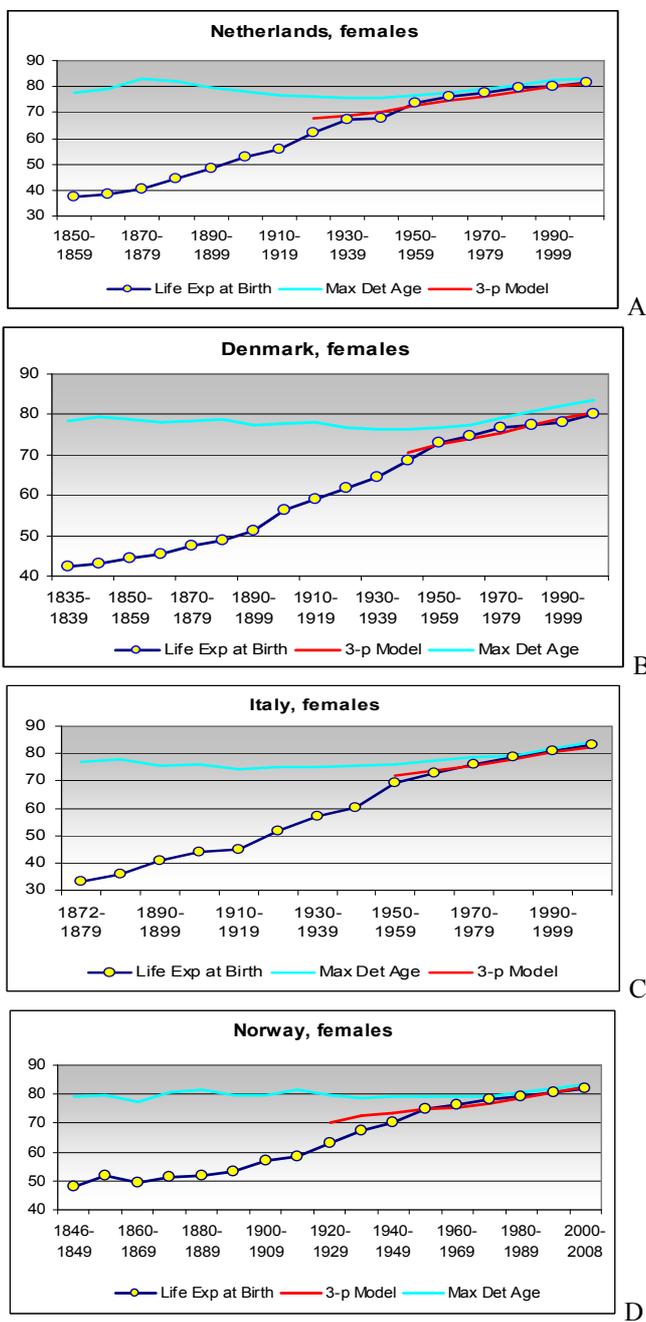

Figure 4. Life expectancy at birth and maximum deterioration age for A) Netherlands, B) Denmark, C) Italy and D) Norway.

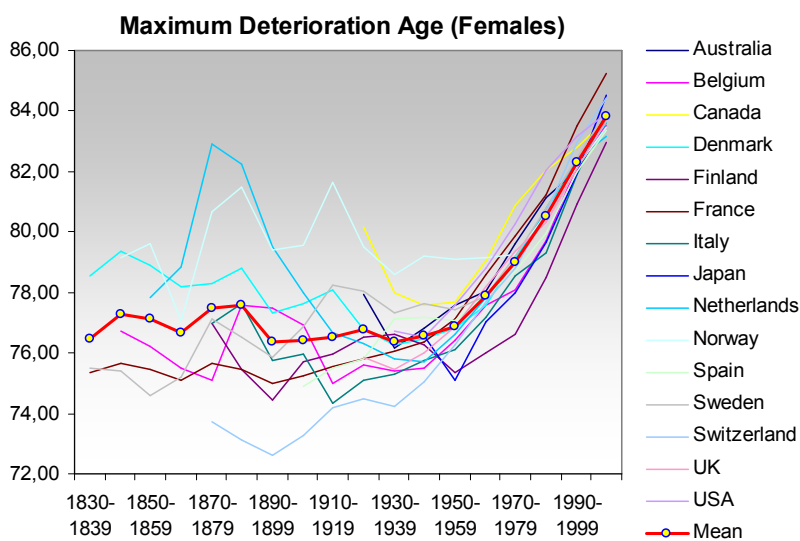

Fig. 5. Maximum deterioration age for 15 countries and mean value.

The age year where the maximum value for the deterioration function is achieved for females in various countries and for several time periods is illustrated in Figure 5. The data for females from various countries for 10 year periods from the human mortality data base are used and the Infant Mortality First Exit Density Model (IM-model) is applied. The maximum deterioration age for females was between 72 – 84 years from 1830 to 1950 for the countries studied (Table II). For all the cases a continuous growth appears for the maximum deterioration age after 1950 until now. That is more important is that the mean value of the maximum deterioration age was between 76 and 78 years for 140 years (1830-1970) irrespective of the fluctuations in the life expectancy supporting the argument for an aging mechanism in the human genes. However, the scientific and medical developments after 1950 gave rise in a gradual increase of the level of the maximum deterioration age from 77 to 84 years in the last 60 years (1950-2010).

## TABLE II

### The Maximum Deterioration Age in 15 Countries (Females)

| Year | Australia | Belgium | Canada | Denmark | Finland | France | Italy | Japan | Netherlands | Norway | Spain | Sweden | Switzerland | UK | USA | Mean |
|---|---|---|---|---|---|---|---|---|---|---|---|---|---|---|---|---|
| 1830-1839 | | | | 78,54 | | 75,36 | | | | | | 75,50 | | | | 76,47 |
| 1840-1849 | | 76,70 | | 79,35 | | 75,67 | | | | 79,15 | | 75,42 | | | | 77,26 |
| 1850-1859 | | 76,20 | | 78,91 | | 75,43 | | | | 79,61 | | 74,61 | | | | 77,10 |
| 1860-1869 | | 75,50 | | 78,18 | | 75,11 | | | 78,86 | 77,07 | | 75,21 | | | | 76,65 |
| 1870-1879 | | 75,10 | | 78,29 | 76,98 | 75,63 | 76,99 | | 82,88 | 80,70 | | 77,12 | 73,70 | | | 77,49 |
| 1880-1889 | | 77,60 | | 78,79 | 75,46 | 75,45 | 77,65 | | 82,25 | 81,48 | | 76,52 | 73,14 | | | 77,59 |
| 1890-1899 | | 77,50 | | 77,32 | 74,44 | 74,98 | 75,76 | | 79,51 | 79,42 | | 75,87 | 72,61 | | | 76,38 |
| 1900-1909 | | 76,90 | | 77,64 | 75,69 | 75,23 | 75,95 | | 77,96 | 79,55 | 74,88 | 76,86 | 73,26 | | | 76,39 |
| 1910-1919 | | 74,99 | | 78,11 | 75,98 | 75,53 | 74,34 | | 76,68 | 81,63 | 75,48 | 78,25 | 74,17 | | | 76,52 |
| 1920-1929 | 77,94 | 75,60 | 80,15 | 76,76 | 76,51 | 75,79 | 75,11 | | 76,34 | 79,52 | 75,78 | 78,01 | 74,49 | 75,87 | | 76,76 |
| 1930-1939 | 76,14 | 75,40 | 78,00 | 78,25 | 76,64 | 76,06 | 75,29 | | 75,81 | 78,61 | 77,11 | 77,34 | 74,21 | 75,47 | 76,71 | 76,36 |
| 1940-1949 | 76,82 | 75,50 | 77,60 | 76,47 | 76,26 | 76,37 | 75,76 | 76,57 | 75,69 | 79,22 | 77,16 | 77,61 | 75,03 | 75,99 | 76,52 | 76,57 |
| 1950-1959 | 77,56 | 76,40 | 77,68 | 76,81 | 75,36 | 77,15 | 76,10 | 75,10 | 76,60 | 79,09 | 77,05 | 77,43 | 76,24 | 76,89 | 77,59 | 76,87 |
| 1960-1969 | 78,04 | 77,60 | 79,03 | 77,57 | 76,02 | 78,61 | 77,15 | 77,00 | 77,84 | 79,14 | 77,92 | 77,87 | 77,71 | 78,20 | 78,82 | 77,90 |
| 1970-1979 | 79,59 | 78,10 | 80,88 | 79,12 | 76,60 | 79,86 | 78,54 | 77,97 | 78,99 | 79,27 | 78,69 | 79,19 | 78,78 | 79,38 | 80,29 | 79,02 |
| 1980-1989 | 81,14 | 79,70 | 82,04 | 80,79 | 78,51 | 81,24 | 79,30 | 79,67 | 80,80 | 80,55 | 79,99 | 80,78 | 80,64 | 80,29 | 82,06 | 80,50 |
| 1990-1999 | 82,07 | 82,20 | 82,79 | 82,22 | 80,95 | 83,51 | 81,88 | 81,93 | 82,38 | 82,16 | 82,00 | 82,96 | 82,96 | 82,10 | 83,13 | 82,32 |
| 2000-2007 | 83,59 | 83,50 | 83,70 | 83,64 | 82,96 | 85,26 | 83,97 | 84,54 | 83,18 | 83,47 | 83,37 | 83,96 | 84,41 | 83,66 | 83,92 | 83,81 |

**Stability of the Deterioration Function Characteristics**

The derivation of the deterioration function of a population allows us to make early estimates for the life expectancy. The main assumption is based on accepting a theory for an internal deterioration mechanism driven by a code which governs the life expectancy. If this assumption holds the deterioration function should include information for the future life expectancy even when using data from periods when the mean life duration was relatively small. In the previous chapter we have found that the maximum value of the deterioration function in many countries was set at high levels even when dealing with mortality data coming from various countries and from the last two centuries.

The next very important point is to estimate the total effect of the deterioration to a population in the course of the life time termed as *DTR*. This is expressed by the following summation formula:

$$DTR = \int_0^t tK(t)dt \approx \sum_0^t tK(t)$$

Where $t$ is the age and $K(t)$ is the deterioration function.

The last formula expresses the expectation that an individual will survive from the deterioration caused in his organism by the deterioration mechanism. The result is given in years of age in a Table like the classical life tables. The deterioration function is estimated from 0 to 117 years a limit set according to the existing death data sets. The estimated life expectancy levels are not the specific levels at the dates of the calculation but refer to future dates when the external influences, illnesses and societal causes will be reduced to a minimum following the advancement of our population status. The life expectancy levels seem to be reached in the recent years in some countries and in the forthcoming decades for others.

DTR will be a strong indicator for the level of life expectancy of a specific population mainly caused by the DNA and genes. Due to its characteristics DTR can also be estimated from only mortality data (number of deaths per age or death distribution) thus making simpler the handling of this indicator even when population data are missing or are not well estimated. Another important point of the last formula is that we can find an estimator of the life expectancy in various age periods and to construct a life table. As it is expected the existence of a deterioration law will result in a population distribution over time thus making possible the construction of life tables by using the population distribution resulting from the deterioration law.

As the introduction of the deterioration function and the DTR indicator are quite new terms introduced when using the stochastic modeling techniques and the first exit time or hitting time theory we have applied the DTR and other forms resulting from the deterioration function to the mortality data for countries included in the Human Mortality Database (HMD). A main advantage by using these data sets is that are systematically collected and developed as to be able to make applications and comparisons between countries. The death data for 10-year periods are preferred as to avoid local fluctuations. However, the results are also strong when using and the other data sets from HMD for 5-year of 1-year periods.

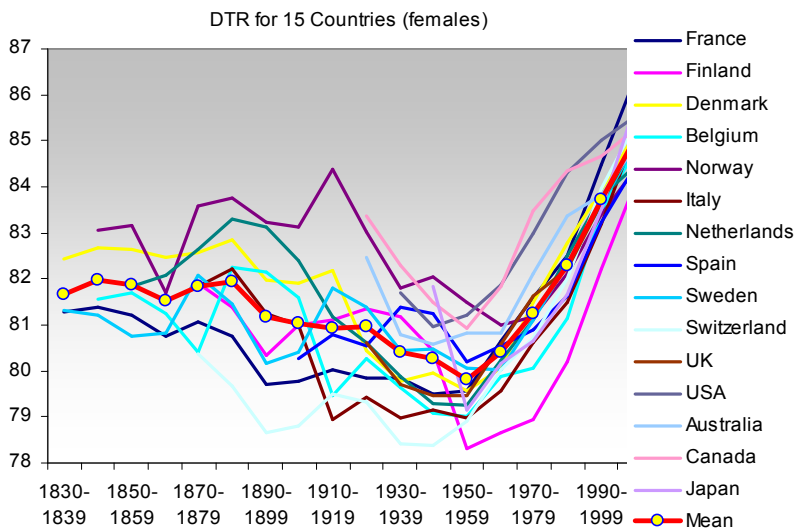

Fig. 6. DTR for 15 countries and mean value.

The DTR is estimated for 15 countries for large time periods. As it is presented in Figure 6 the DTR for 150 years from 1830 to 1980 was between 78 and 84 years of age with the mean value to be from 79.80 to 81.98 years (see Table III). The lowest value 79.80 years was achieved the period 1950-1959. The main conclusion is that the DTR can be used as a measure of the future life expectancy levels. An example is based on the estimates for Sweden from 1751. The same method can be applied for other countries.

TABLE III

The DTR effect in 15 Countries (Females)

| Year | Australia | Belgium | Canada | Denmark | Finland | France | Italy | Japan | Netherlands | Norway | Spain | Sweden | Switzerland | UK | USA | Mean |
|---|---|---|---|---|---|---|---|---|---|---|---|---|---|---|---|---|
| 1830-1839 | | | | 82,43 | | 81,29 | | | | | | 81,30 | | | | 81,67 |
| 1840-1849 | | 81,55 | | 82,68 | | 81,39 | | | | 83,07 | | 81,20 | | | | 81,98 |
| 1850-1859 | | 81,70 | | 82,65 | | 81,20 | | | 81,84 | 83,16 | | 80,76 | | | | 81,89 |
| 1860-1869 | | 81,23 | | 82,47 | | 80,75 | | | 82,07 | 81,71 | | 80,81 | | | | 81,51 |
| 1870-1879 | | 80,41 | | 82,57 | 81,92 | 81,07 | 81,85 | | 82,62 | 83,59 | | 82,07 | 80,37 | | | 81,83 |
| 1880-1889 | | 82,25 | | 82,86 | 81,37 | 80,76 | 82,22 | | 83,30 | 83,77 | | 81,44 | 79,66 | | | 81,96 |
| 1890-1899 | | 82,16 | | 81,98 | 80,34 | 79,70 | 81,23 | | 83,12 | 83,22 | | 80,15 | 78,67 | | | 81,17 |
| 1900-1909 | | 81,61 | | 81,89 | 81,01 | 79,77 | 80,98 | | 82,38 | 83,13 | 80,26 | 80,42 | 78,82 | | | 81,03 |
| 1910-1919 | | 79,45 | | 82,20 | 81,10 | 80,01 | 78,93 | | 81,18 | 84,39 | 80,78 | 81,81 | 79,49 | | | 80,93 |
| 1920-1929 | 82,47 | 80,28 | | 80,43 | 81,35 | 79,86 | 79,42 | | 80,63 | 83,02 | 80,54 | 81,37 | 79,32 | 80,61 | | 80,98 |
| 1930-1939 | 80,80 | 79,64 | 82,30 | 79,77 | 81,16 | 79,86 | 78,98 | | 79,88 | 81,80 | 81,40 | 80,45 | 78,42 | 79,72 | 81,70 | 80,42 |
| 1940-1949 | 80,58 | 79,09 | 81,48 | 79,94 | 80,46 | 79,50 | 79,16 | 81,85 | 79,28 | 82,05 | 81,25 | 80,48 | 78,39 | 79,46 | 80,97 | 80,26 |
| 1950-1959 | 80,82 | 79,02 | 80,92 | 79,57 | 78,30 | 79,59 | 78,98 | 79,16 | 79,26 | 81,48 | 80,20 | 80,06 | 78,91 | 79,48 | 81,21 | 79,80 |
| 1960-1969 | 80,81 | 79,90 | 81,82 | 80,05 | 78,66 | 80,66 | 79,58 | 80,15 | 80,22 | 81,01 | 80,55 | 80,02 | 80,02 | 80,58 | 81,89 | 80,39 |
| 1970-1979 | 82,13 | 80,07 | 83,49 | 81,47 | 78,95 | 81,59 | 80,60 | 80,64 | 81,10 | 81,18 | 80,89 | 81,12 | 80,72 | 81,63 | 82,98 | 81,24 |
| 1980-1989 | 83,37 | 81,15 | 84,33 | 82,77 | 80,21 | 82,49 | 81,49 | 81,67 | 82,51 | 82,12 | 81,64 | 82,31 | 82,00 | 82,21 | 84,32 | 82,31 |
| 1990-1999 | 83,82 | 83,47 | 84,68 | 83,91 | 82,21 | 84,49 | 83,18 | 83,52 | 83,80 | 83,41 | 83,25 | 83,61 | 83,97 | 83,66 | 85,01 | 83,73 |
| 2000-2007 | 84,87 | 84,79 | 85,24 | 85,14 | 84,07 | 86,36 | 85,15 | 85,78 | 84,46 | 84,39 | 84,45 | 84,89 | 85,36 | 84,99 | 85,53 | 85,03 |

The next Figure 7 illustrates the graphs from the DTR application to Sweden (females) for the period 1751-1759 (red line) and 1950-1959 (brown line). The bleu line expresses the life expectancy estimates for 2000-2008 downloaded from the human mortality database. Our application includes the periods 1800-1809, 1850-1859 and 2000-2008. The results of the DTR method refer to the future limits of the life expectancy in connection to the limits of the human organism and the developments of science. Our results for 2000-2008 suggest a future level for the life expectancy at birth in Sweden at 84,89 years, 2,34 years higher than the period 2000-2008 (see Table V).

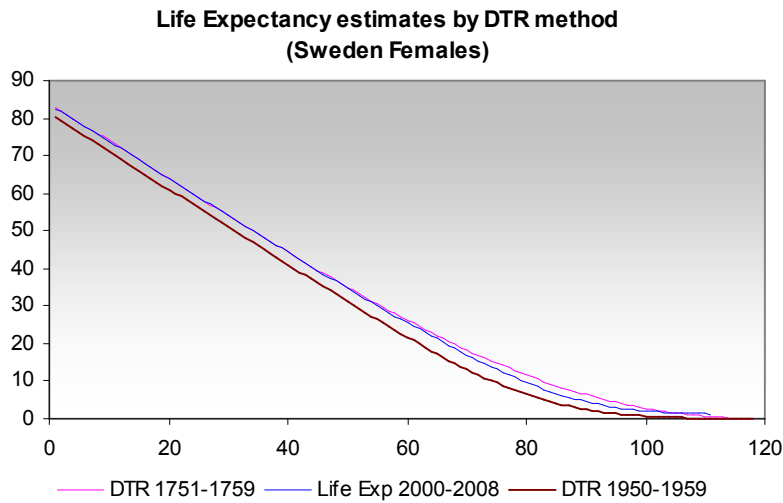

Fig. 7. DTR application to Sweden (females) for the period 1751-1759 (red line) and 1950-1959 (brown line). Life expectancy estimates for 2000-2008 are provided from the human mortality database (bleu line).

**Estimation of the life expectancy by the DTR system**

The estimation of the life expectancy by using the DTR system is presented in Table IV. In the first column the data for the deaths per age are presented. In our example the data for the period 1751-1759 for Sweden (females) are included. In the second column the data are normalized divided by their sum. This is important in order to compare data sets from various periods and from several countries. The next step is to apply the first exit time model including infant mortality (IM-model) to the normalized data. The fitting estimates appear in the third column. The fourth column is the age in years. The next step is to estimate the Deterioration Function K(x). We need the values for the parameter $b$ and the exponent $c$ already estimated in the previous step when applying the IM-model. The deterioration function values are normalized and included in the sixth column. Then the DTR values are estimated in the seventh column. These values are computed as a multiplication: $xK(x)$ and then are normalized and stored in the eight column. Then the DTR survival curve is estimated and included in column nine. We apply the following formula for the survival curve (SC):

$$SC_x = 1 - \sum_0^{x-1} xK(x).$$

The final step is the formulation of the tenth column including the DTR life expectancy list according to age from the survival curve. The life expectancy $e_x$ at age $x$ is calculated by: $e_x = \sum_0^n SC_x - \sum_0^x SC_x$, where $n$ was selected the age 117 as a level according to our experience. However, the selection of a different age level will change the estimates for the life expectancy. The best approach will be the selection of a level age which is in accordance to reality.

The results of the application of the DTR method for various time periods in Sweden for females are summarized in Table V. The estimated life expectancy values are very close to that achieved in recent years as is also illustrated in Figure 8 thus supporting our argument for using the DTR method to forecasts the future trends of life expectancy. Even from the period 1751-1759 we can estimate a life expectancy at birth at the level of 82.92 years for Sweden, females. This value is very close to that of the period 2000-2008 (82.55 years).

| | | | | | | | | | |
|---|---|---|---|---|---|---|---|---|---|
| **TABLE IV** | | | | | | | | | |
| **DTR System (Sweden, Females)** | | | | | | | | | |
| 1751-1759 Data | Data Normalised | IM Model Estimates | Year x | K(x) | K(x) Normalised | xK(x) | xK(x) Normalised | Survival curve | Life Expectancy |
| 57425 | 0,2473 | 0,2317 | 0 | 4,56E-05 | 6,004E-05 | 0 | 0 | 1 | 82,92 |
| 14438 | 0,0622 | 0,0873 | 1 | 0,000112 | 0,0001478 | 0,000148 | 2,044E-06 | 1 | 81,92 |
| 9467 | 0,0408 | 0,0486 | 2 | 0,00019 | 0,0002505 | 0,000501 | 6,926E-06 | 1 | 80,92 |
| 6739 | 0,0290 | 0,0319 | 3 | 0,000276 | 0,000364 | 0,001092 | 1,51E-05 | 0,99999 | 79,92 |
| 4918 | 0,0212 | 0,0231 | 4 | 0,000369 | 0,0004866 | 0,001946 | 2,691E-05 | 0,99998 | 78,92 |
| 3619 | 0,0156 | 0,0177 | 5 | 0,000468 | 0,0006167 | 0,003083 | 4,264E-05 | 0,99995 | 77,92 |
| 2623 | 0,0113 | 0,0142 | 6 | 0,000572 | 0,0007535 | 0,004521 | 6,252E-05 | 0,99991 | 76,92 |
| 1875 | 0,0081 | 0,0117 | 7 | 0,00068 | 0,0008964 | 0,006275 | 8,676E-05 | 0,99984 | 75,92 |
| 1375 | 0,0059 | 0,0100 | 8 | 0,000793 | 0,0010447 | 0,008357 | 0,0001156 | 0,99976 | 74,92 |
| 1124 | 0,0048 | 0,0087 | 9 | 0,000909 | 0,001198 | 0,010782 | 0,0001491 | 0,99964 | 73,92 |
| 1078 | 0,0046 | 0,0077 | 10 | 0,001029 | 0,001356 | 0,01356 | 0,0001875 | 0,99949 | 72,92 |
| 1065 | 0,0046 | 0,0069 | 11 | 0,001152 | 0,0015184 | 0,016702 | 0,000231 | 0,9993 | 71,92 |
| 1044 | 0,0045 | 0,0063 | 12 | 0,001279 | 0,0016849 | 0,020218 | 0,0002796 | 0,99907 | 70,92 |
| 1013 | 0,0044 | 0,0058 | 13 | 0,001408 | 0,0018552 | 0,024117 | 0,0003335 | 0,99879 | 69,92 |
| 973 | 0,0042 | 0,0054 | 14 | 0,00154 | 0,0020292 | 0,028409 | 0,0003928 | 0,99846 | 68,92 |
| 929 | 0,0040 | 0,0051 | 15 | 0,001675 | 0,0022067 | 0,0331 | 0,0004577 | 0,99807 | 67,93 |
| 899 | 0,0039 | 0,0049 | 16 | 0,001812 | 0,0023875 | 0,038199 | 0,0005282 | 0,99761 | 66,93 |
| 888 | 0,0038 | 0,0047 | 17 | 0,001952 | 0,0025714 | 0,043714 | 0,0006045 | 0,99708 | 65,93 |
| 897 | 0,0039 | 0,0046 | 18 | 0,002093 | 0,0027584 | 0,049651 | 0,0006865 | 0,99648 | 64,93 |
| 926 | 0,0040 | 0,0045 | 19 | 0,002237 | 0,0029482 | 0,056016 | 0,0007746 | 0,99579 | 63,94 |
| 969 | 0,0042 | 0,0044 | 20 | 0,002384 | 0,0031408 | 0,062817 | 0,0008686 | 0,99502 | 62,94 |
| 1011 | 0,0044 | 0,0044 | 21 | 0,002532 | 0,0033361 | 0,070058 | 0,0009687 | 0,99415 | 61,95 |
| 1046 | 0,0045 | 0,0044 | 22 | 0,002682 | 0,0035338 | 0,077744 | 0,001075 | 0,99318 | 60,95 |
| 1076 | 0,0046 | 0,0044 | 23 | 0,002834 | 0,003734 | 0,085882 | 0,0011875 | 0,9921 | 59,96 |
| 1100 | 0,0047 | 0,0045 | 24 | 0,002987 | 0,0039364 | 0,094474 | 0,0013063 | 0,99092 | 58,97 |
| 1122 | 0,0048 | 0,0045 | 25 | 0,003143 | 0,0041411 | 0,103526 | 0,0014315 | 0,98961 | 57,97 |
| 1158 | 0,0050 | 0,0046 | 26 | 0,0033 | 0,0043477 | 0,113041 | 0,0015631 | 0,98818 | 56,99 |
| 1214 | 0,0052 | 0,0047 | 27 | 0,003458 | 0,0045563 | 0,123021 | 0,0017011 | 0,98662 | 56,00 |
| 1288 | 0,0055 | 0,0048 | 28 | 0,003618 | 0,0047668 | 0,13347 | 0,0018456 | 0,98491 | 55,01 |
| 1382 | 0,0060 | 0,0049 | 29 | 0,003779 | 0,0049789 | 0,144389 | 0,0019965 | 0,98307 | 54,03 |
| 1482 | 0,0064 | 0,0050 | 30 | 0,003941 | 0,0051927 | 0,15578 | 0,002154 | 0,98107 | 53,04 |
| 1537 | 0,0066 | 0,0051 | 31 | 0,004104 | 0,0054078 | 0,167643 | 0,0023181 | 0,97892 | 52,06 |
| 1535 | 0,0066 | 0,0053 | 32 | 0,004268 | 0,0056243 | 0,179979 | 0,0024887 | 0,9766 | 51,08 |
| 1476 | 0,0064 | 0,0054 | 33 | 0,004433 | 0,005842 | 0,192787 | 0,0026658 | 0,97411 | 50,11 |
| 1359 | 0,0059 | 0,0056 | 34 | 0,0046 | 0,0060608 | 0,206066 | 0,0028494 | 0,97145 | 49,13 |
| 1207 | 0,0052 | 0,0058 | 35 | 0,004766 | 0,0062804 | 0,219813 | 0,0030395 | 0,9686 | 48,16 |
| 1103 | 0,0047 | 0,0059 | 36 | 0,004934 | 0,0065007 | 0,234027 | 0,003236 | 0,96556 | 47,19 |
| 1068 | 0,0046 | 0,0061 | 37 | 0,005101 | 0,0067217 | 0,248702 | 0,0034389 | 0,96232 | 46,23 |
| 1104 | 0,0048 | 0,0063 | 38 | 0,005269 | 0,006943 | 0,263834 | 0,0036482 | 0,95888 | 45,26 |
| 1209 | 0,0052 | 0,0065 | 39 | 0,005437 | 0,0071645 | 0,279417 | 0,0038636 | 0,95523 | 44,31 |
| 1362 | 0,0059 | 0,0067 | 40 | 0,005605 | 0,0073861 | 0,295444 | 0,0040853 | 0,95137 | 43,35 |
| 1469 | 0,0063 | 0,0069 | 41 | 0,005774 | 0,0076075 | 0,311908 | 0,0043129 | 0,94729 | 42,40 |
| 1509 | 0,0065 | 0,0071 | 42 | 0,005941 | 0,0078286 | 0,328799 | 0,0045465 | 0,94297 | 41,45 |
| 1481 | 0,0064 | 0,0073 | 43 | 0,006109 | 0,008049 | 0,346107 | 0,0047858 | 0,93843 | 40,51 |
| 1386 | 0,0060 | 0,0076 | 44 | 0,006275 | 0,0082686 | 0,36382 | 0,0050307 | 0,93364 | 39,57 |
| 1245 | 0,0054 | 0,0078 | 45 | 0,006441 | 0,0084872 | 0,381925 | 0,0052811 | 0,92861 | 38,64 |
| 1144 | 0,0049 | 0,0080 | 46 | 0,006606 | 0,0087045 | 0,400408 | 0,0055367 | 0,92333 | 37,71 |
| 1106 | 0,0048 | 0,0082 | 47 | 0,00677 | 0,0089203 | 0,419253 | 0,0057972 | 0,91779 | 36,78 |
| 1129 | 0,0049 | 0,0084 | 48 | 0,006932 | 0,0091342 | 0,438443 | 0,0060626 | 0,91199 | 35,87 |
| 1215 | 0,0052 | 0,0087 | 49 | 0,007093 | 0,0093461 | 0,457959 | 0,0063324 | 0,90593 | 34,95 |
| 1346 | 0,0058 | 0,0089 | 50 | 0,007252 | 0,0095556 | 0,47778 | 0,0066065 | 0,8996 | 34,05 |
| 1450 | 0,0062 | 0,0091 | 51 | 0,007409 | 0,0097625 | 0,497886 | 0,0068845 | 0,89299 | 33,15 |
| 1509 | 0,0065 | 0,0093 | 52 | 0,007564 | 0,0099664 | 0,518253 | 0,0071662 | 0,88611 | 32,26 |
| 1524 | 0,0066 | 0,0095 | 53 | 0,007716 | 0,0101671 | 0,538855 | 0,007451 | 0,87894 | 31,37 |
| 1494 | 0,0064 | 0,0097 | 54 | 0,007866 | 0,0103642 | 0,559667 | 0,0077388 | 0,87149 | 30,49 |
| 1438 | 0,0062 | 0,0099 | 55 | 0,008012 | 0,0105575 | 0,58066 | 0,0080291 | 0,86375 | 29,62 |
| 1431 | 0,0062 | 0,0101 | 56 | 0,008156 | 0,0107465 | 0,601805 | 0,0083215 | 0,85572 | 28,76 |
| 1491 | 0,0064 | 0,0102 | 57 | 0,008296 | 0,010931 | 0,62307 | 0,0086155 | 0,8474 | 27,90 |
| 1619 | 0,0070 | 0,0104 | 58 | 0,008432 | 0,0111107 | 0,644423 | 0,0089108 | 0,83879 | 27,05 |
| 1814 | 0,0078 | 0,0106 | 59 | 0,008565 | 0,0112852 | 0,66583 | 0,0092068 | 0,82988 | 26,21 |
| 2053 | 0,0088 | 0,0107 | 60 | 0,008693 | 0,0114543 | 0,687255 | 0,009503 | 0,82067 | 25,38 |
| 2246 | 0,0097 | 0,0108 | 61 | 0,008817 | 0,0116174 | 0,708662 | 0,0097991 | 0,81117 | 24,56 |
| 2371 | 0,0102 | 0,0109 | 62 | 0,008936 | 0,0117744 | 0,730013 | 0,0100943 | 0,80137 | 23,75 |
| 2428 | 0,0105 | 0,0110 | 63 | 0,00905 | 0,0119249 | 0,751269 | 0,0103882 | 0,79127 | 22,95 |
| 2416 | 0,0104 | 0,0111 | 64 | 0,009159 | 0,0120686 | 0,77239 | 0,0106803 | 0,78088 | 22,16 |
| 2357 | 0,0102 | 0,0111 | 65 | 0,009263 | 0,0122052 | 0,793336 | 0,0109699 | 0,7702 | 21,38 |
| 2343 | 0,0101 | 0,0112 | 66 | 0,009361 | 0,0123343 | 0,814064 | 0,0112565 | 0,75923 | 20,61 |
| 2395 | 0,0103 | 0,0112 | 67 | 0,009453 | 0,0124557 | 0,834534 | 0,0115395 | 0,74798 | 19,85 |
| 2512 | 0,0108 | 0,0111 | 68 | 0,009539 | 0,0125692 | 0,854702 | 0,0118184 | 0,73644 | 19,10 |
| 2696 | 0,0116 | 0,0111 | 69 | 0,009619 | 0,0126743 | 0,874527 | 0,0120926 | 0,72462 | 18,36 |
| 2918 | 0,0126 | 0,0110 | 70 | 0,009692 | 0,0127709 | 0,893965 | 0,0123613 | 0,71253 | 17,64 |
| 3060 | 0,0132 | 0,0109 | 71 | 0,009759 | 0,0128588 | 0,912975 | 0,0126242 | 0,70017 | 16,93 |
| 3096 | 0,0133 | 0,0108 | 72 | 0,009819 | 0,0129377 | 0,931515 | 0,0128806 | 0,68754 | 16,23 |
| 3025 | 0,0130 | 0,0107 | 73 | 0,009872 | 0,0130074 | 0,949543 | 0,0131298 | 0,67466 | 15,54 |
| 2846 | 0,0123 | 0,0105 | 74 | 0,009917 | 0,0130678 | 0,967019 | 0,0133715 | 0,66153 | 14,87 |
| 2588 | 0,0111 | 0,0103 | 75 | 0,009956 | 0,0131187 | 0,983905 | 0,013605 | 0,64816 | 14,20 |
| 2358 | 0,0102 | 0,0101 | 76 | 0,009987 | 0,01316 | 1,000161 | 0,0138298 | 0,63455 | 13,56 |
| 2184 | 0,0094 | 0,0098 | 77 | 0,010011 | 0,0131916 | 1,015751 | 0,0140453 | 0,62072 | 12,92 |
| 2065 | 0,0089 | 0,0096 | 78 | 0,010028 | 0,0132133 | 1,030641 | 0,0142512 | 0,60668 | 12,30 |
| 2003 | 0,0086 | 0,0093 | 79 | 0,010037 | 0,0132253 | 1,044797 | 0,014447 | 0,59243 | 11,69 |
| 1979 | 0,0085 | 0,0089 | 80 | 0,010039 | 0,0132273 | 1,058188 | 0,0146321 | 0,57798 | 11,10 |
| 1927 | 0,0083 | 0,0086 | 81 | 0,010033 | 0,0132196 | 1,070785 | 0,0148063 | 0,56335 | 10,52 |
| 1828 | 0,0079 | 0,0082 | 82 | 0,010019 | 0,013202 | 1,082563 | 0,0149692 | 0,54854 | 9,96 |
| 1683 | 0,0072 | 0,0079 | 83 | 0,009999 | 0,0131747 | 1,093498 | 0,0151204 | 0,53357 | 9,41 |
| 1492 | 0,0064 | 0,0075 | 84 | 0,009971 | 0,0131377 | 1,103567 | 0,0152596 | 0,51845 | 8,88 |
| 1267 | 0,0055 | 0,0071 | 85 | 0,009935 | 0,0130912 | 1,112754 | 0,0153866 | 0,50319 | 8,36 |
| 1061 | 0,0046 | 0,0067 | 86 | 0,009893 | 0,0130354 | 1,121042 | 0,0155013 | 0,48781 | 7,86 |
| 887 | 0,0038 | 0,0062 | 87 | 0,009844 | 0,0129703 | 1,12842 | 0,0156033 | 0,47231 | 7,37 |
| 743 | 0,0032 | 0,0058 | 88 | 0,009787 | 0,0128963 | 1,134876 | 0,0156925 | 0,4567 | 6,90 |
| 631 | 0,0027 | 0,0054 | 89 | 0,009725 | 0,0128136 | 1,140407 | 0,015769 | 0,44101 | 6,44 |
| 586 | 0,0025 | 0,0050 | 90 | 0,009655 | 0,0127223 | 1,145007 | 0,0158326 | 0,42524 | 6,00 |
| 484 | 0,0021 | 0,0046 | 91 | 0,00958 | 0,0126228 | 1,148676 | 0,0158834 | 0,40941 | 5,57 |
| 395 | 0,0017 | 0,0042 | 92 | 0,009498 | 0,0125154 | 1,151418 | 0,0159213 | 0,39352 | 5,16 |
| 318 | 0,0014 | 0,0038 | 93 | 0,009411 | 0,0124004 | 1,153238 | 0,0159464 | 0,3776 | 4,77 |
| 252 | 0,0011 | 0,0034 | 94 | 0,009318 | 0,0122781 | 1,154144 | 0,015959 | 0,36166 | 4,39 |
| 197 | 0,0008 | 0,0031 | 95 | 0,00922 | 0,0121489 | 1,154147 | 0,015959 | 0,3457 | 4,03 |
| 151 | 0,0007 | 0,0028 | 96 | 0,009117 | 0,0120131 | 1,153262 | 0,0159468 | 0,32974 | 3,68 |
| 115 | 0,0005 | 0,0024 | 97 | 0,009009 | 0,0118712 | 1,151505 | 0,0159225 | 0,31379 | 3,36 |
| 86 | 0,0004 | 0,0021 | 98 | 0,008897 | 0,0117234 | 1,148894 | 0,0158864 | 0,29787 | 3,04 |
| 63 | 0,0003 | 0,0019 | 99 | 0,008781 | 0,0115702 | 1,145451 | 0,0158388 | 0,28198 | 2,74 |
| 45 | 0,0002 | 0,0016 | 100 | 0,008661 | 0,011412 | 1,141199 | 0,01578 | 0,26614 | 2,46 |
| 32 | 0,0001 | 0,0014 | 101 | 0,008537 | 0,0112491 | 1,136163 | 0,0157103 | 0,25036 | 2,20 |
| 23 | 0,0001 | 0,0012 | 102 | 0,00841 | 0,0110821 | 1,13037 | 0,0156302 | 0,23465 | 1,95 |
| 15 | 0,0001 | 0,0010 | 103 | 0,008281 | 0,0109111 | 1,123847 | 0,01554 | 0,21902 | 1,71 |
| 10 | 0,0000 | 0,0008 | 104 | 0,008148 | 0,0107368 | 1,116625 | 0,0154402 | 0,20348 | 1,49 |
| 6 | 0,0000 | 0,0007 | 105 | 0,008014 | 0,0105594 | 1,108734 | 0,0153311 | 0,18804 | 1,29 |
| 3 | 0,0000 | 0,0006 | 106 | 0,007877 | 0,0103793 | 1,100205 | 0,0152131 | 0,17271 | 1,10 |
| 1 | 0,0000 | 0,0005 | 107 | 0,007739 | 0,0101969 | 1,091072 | 0,0150868 | 0,1575 | 0,93 |
| 0 | 0,0000 | 0,0004 | 108 | 0,007599 | 0,0100127 | 1,081367 | 0,0149526 | 0,14241 | 0,77 |
| 0 | 0,0000 | 0,0003 | 109 | 0,007458 | 0,0098268 | 1,071124 | 0,014811 | 0,12746 | 0,63 |
| 0 | 0,0000 | 0,0002 | 110 | 0,007316 | 0,0096398 | 1,060376 | 0,0146624 | 0,11265 | 0,50 |
| | 0,0000 | 0,0002 | 111 | 0,007173 | 0,0094519 | 1,049157 | 0,0145073 | 0,09799 | 0,39 |
| | 0,0000 | 0,0001 | 112 | 0,00703 | 0,0092634 | 1,037502 | 0,0143461 | 0,08348 | 0,29 |
| | 0,0000 | 0,0001 | 113 | 0,006887 | 0,0090747 | 1,025442 | 0,0141793 | 0,06913 | 0,21 |
| | 0,0000 | 0,0001 | 114 | 0,006744 | 0,0088861 | 1,013012 | 0,0140075 | 0,05495 | 0,14 |
| | 0,0000 | 0,0001 | 115 | 0,006601 | 0,0086978 | 1,000245 | 0,0138309 | 0,04095 | 0,08 |
| | 0,0000 | 0,0000 | 116 | 0,006459 | 0,0085101 | 0,987171 | 0,0136501 | 0,02712 | 0,04 |
| | 0,0000 | 0,0000 | 117 | 0,006317 | 0,0083233 | 0,973824 | 0,0134656 | 0,01347 | 0,01 |

**TABLE V**

Results from the DTR System for the Future Life Expectancy in Sweden (Females)

| Year x | 1751-1759 | 1800-1809 | 1850-1859 | 1900-1909 | 1950-1959 | 2000-2008 | Life Exp 2000-2008 |
|---|---|---|---|---|---|---|---|
| 0 | 82,92 | 81,21 | 80,76 | 80,42 | 80,06 | 84,89 | 82,55 |
| 1 | 81,92 | 80,21 | 79,76 | 79,42 | 79,06 | 83,89 | 81,77 |
| 2 | 80,92 | 79,21 | 78,76 | 78,42 | 78,06 | 82,89 | 80,79 |
| 3 | 79,92 | 78,21 | 77,76 | 77,42 | 77,06 | 81,89 | 79,8 |
| 4 | 78,92 | 77,21 | 76,76 | 76,42 | 76,06 | 80,89 | 78,81 |
| 5 | 77,92 | 76,21 | 75,76 | 75,42 | 75,06 | 79,89 | 77,82 |
| 6 | 76,92 | 75,21 | 74,76 | 74,42 | 74,06 | 78,89 | 76,83 |
| 7 | 75,92 | 74,21 | 73,76 | 73,42 | 73,06 | 77,89 | 75,84 |
| 8 | 74,92 | 73,21 | 72,76 | 72,42 | 72,06 | 76,89 | 74,84 |
| 9 | 73,92 | 72,21 | 71,76 | 71,42 | 71,06 | 75,89 | 73,85 |
| 10 | 72,92 | 71,21 | 70,76 | 70,42 | 70,06 | 74,89 | 72,85 |
| 11 | 71,92 | 70,21 | 69,76 | 69,42 | 69,06 | 73,89 | 71,86 |
| 12 | 70,92 | 69,21 | 68,76 | 68,42 | 68,06 | 72,89 | 70,86 |
| 13 | 69,92 | 68,21 | 67,76 | 67,42 | 67,06 | 71,89 | 69,87 |
| 14 | 68,92 | 67,22 | 66,76 | 66,42 | 66,06 | 70,89 | 68,88 |
| 15 | 67,93 | 66,22 | 65,76 | 65,42 | 65,06 | 69,89 | 67,88 |
| 16 | 66,93 | 65,22 | 64,76 | 64,42 | 64,06 | 68,89 | 66,9 |
| 17 | 65,93 | 64,22 | 63,77 | 63,42 | 63,06 | 67,89 | 65,91 |
| 18 | 64,93 | 63,22 | 62,77 | 62,42 | 62,06 | 66,89 | 64,92 |
| 19 | 63,94 | 62,22 | 61,77 | 61,42 | 61,06 | 65,89 | 63,94 |
| 20 | 62,94 | 61,23 | 60,77 | 60,42 | 60,06 | 64,89 | 62,96 |
| 21 | 61,95 | 60,23 | 59,77 | 59,42 | 59,06 | 63,89 | 61,97 |
| 22 | 60,95 | 59,24 | 58,78 | 58,42 | 58,07 | 62,89 | 60,99 |
| 23 | 59,96 | 58,24 | 57,78 | 57,42 | 57,07 | 61,89 | 60 |
| 24 | 58,97 | 57,25 | 56,79 | 56,42 | 56,07 | 60,89 | 59,02 |
| 25 | 57,97 | 56,25 | 55,79 | 55,42 | 55,07 | 59,89 | 58,03 |
| 26 | 56,99 | 55,26 | 54,80 | 54,42 | 54,07 | 58,89 | 57,05 |
| 27 | 56,00 | 54,27 | 53,81 | 53,43 | 53,07 | 57,89 | 56,06 |
| 28 | 55,01 | 53,28 | 52,82 | 52,43 | 52,07 | 56,89 | 55,08 |
| 29 | 54,03 | 52,30 | 51,83 | 51,43 | 51,07 | 55,89 | 54,09 |
| 30 | 53,04 | 51,31 | 50,84 | 50,43 | 50,07 | 54,89 | 53,11 |
| 31 | 52,06 | 50,33 | 49,86 | 49,43 | 49,07 | 53,89 | 52,12 |
| 32 | 51,08 | 49,35 | 48,87 | 48,44 | 48,07 | 52,89 | 51,14 |
| 33 | 50,11 | 48,37 | 47,89 | 47,44 | 47,07 | 51,89 | 50,16 |
| 34 | 49,13 | 47,39 | 46,91 | 46,44 | 46,07 | 50,89 | 49,17 |
| 35 | 48,16 | 46,41 | 45,93 | 45,45 | 45,08 | 49,89 | 48,19 |
| 36 | 47,19 | 45,44 | 44,96 | 44,46 | 44,08 | 48,89 | 47,21 |
| 37 | 46,23 | 44,47 | 43,98 | 43,46 | 43,08 | 47,89 | 46,24 |
| 38 | 45,26 | 43,51 | 43,02 | 42,47 | 42,09 | 46,89 | 45,26 |
| 39 | 44,31 | 42,55 | 42,05 | 41,48 | 41,09 | 45,89 | 44,29 |
| 40 | 43,35 | 41,59 | 41,09 | 40,49 | 40,10 | 44,89 | 43,31 |
| 41 | 42,40 | 40,63 | 40,13 | 39,51 | 39,11 | 43,89 | 42,34 |
| 42 | 41,45 | 39,68 | 39,17 | 38,52 | 38,11 | 42,89 | 41,37 |
| 43 | 40,51 | 38,74 | 38,22 | 37,54 | 37,12 | 41,89 | 40,41 |
| 44 | 39,57 | 37,79 | 37,28 | 36,56 | 36,13 | 40,89 | 39,45 |
| 45 | 38,64 | 36,86 | 36,34 | 35,58 | 35,15 | 39,89 | 38,49 |
| 46 | 37,71 | 35,93 | 35,40 | 34,60 | 34,16 | 38,89 | 37,53 |
| 47 | 36,78 | 35,00 | 34,47 | 33,63 | 33,18 | 37,90 | 36,58 |
| 48 | 35,87 | 34,08 | 33,55 | 32,66 | 32,20 | 36,90 | 35,64 |
| 49 | 34,95 | 33,17 | 32,63 | 31,70 | 31,22 | 35,90 | 34,7 |
| 50 | 34,05 | 32,26 | 31,72 | 30,74 | 30,25 | 34,91 | 33,76 |
| 51 | 33,15 | 31,36 | 30,81 | 29,78 | 29,28 | 33,91 | 32,83 |
| 52 | 32,26 | 30,47 | 29,91 | 28,83 | 28,31 | 32,92 | 31,91 |
| 53 | 31,37 | 29,58 | 29,02 | 27,89 | 27,35 | 31,92 | 30,98 |
| 54 | 30,49 | 28,71 | 28,14 | 26,95 | 26,39 | 30,93 | 30,07 |
| 55 | 29,62 | 27,84 | 27,27 | 26,02 | 25,44 | 29,94 | 29,15 |
| 56 | 28,76 | 26,98 | 26,40 | 25,10 | 24,50 | 28,95 | 28,24 |
| 57 | 27,90 | 26,12 | 25,55 | 24,18 | 23,56 | 27,97 | 27,35 |
| 58 | 27,05 | 25,28 | 24,70 | 23,27 | 22,63 | 26,98 | 26,46 |
| 59 | 26,21 | 24,45 | 23,86 | 22,37 | 21,71 | 26,00 | 25,57 |
| 60 | 25,38 | 23,62 | 23,04 | 21,48 | 20,79 | 25,02 | 24,7 |

| 61  | 24,56 | 22,81 | 22,22 | 20,61 | 19,89 | 24,05 | 23,82 |
| 62  | 23,75 | 22,01 | 21,42 | 19,74 | 19,00 | 23,07 | 22,96 |
| 63  | 22,95 | 21,22 | 20,63 | 18,88 | 18,11 | 22,11 | 22,1  |
| 64  | 22,16 | 20,44 | 19,85 | 18,04 | 17,25 | 21,15 | 21,24 |
| 65  | 21,38 | 19,67 | 19,08 | 17,21 | 16,39 | 20,19 | 20,4  |
| 66  | 20,61 | 18,91 | 18,32 | 16,40 | 15,55 | 19,24 | 19,57 |
| 67  | 19,85 | 18,17 | 17,58 | 15,60 | 14,73 | 18,30 | 18,74 |
| 68  | 19,10 | 17,44 | 16,85 | 14,82 | 13,92 | 17,37 | 17,92 |
| 69  | 18,36 | 16,72 | 16,13 | 14,06 | 13,13 | 16,45 | 17,11 |
| 70  | 17,64 | 16,02 | 15,43 | 13,31 | 12,36 | 15,54 | 16,32 |
| 71  | 16,93 | 15,33 | 14,74 | 12,58 | 11,61 | 14,65 | 15,53 |
| 72  | 16,23 | 14,65 | 14,07 | 11,87 | 10,88 | 13,76 | 14,76 |
| 73  | 15,54 | 13,99 | 13,41 | 11,18 | 10,18 | 12,89 | 14    |
| 74  | 14,87 | 13,34 | 12,77 | 10,52 | 9,50  | 12,04 | 13,24 |
| 75  | 14,20 | 12,70 | 12,15 | 9,87  | 8,84  | 11,21 | 12,51 |
| 76  | 13,56 | 12,09 | 11,53 | 9,25  | 8,21  | 10,40 | 11,78 |
| 77  | 12,92 | 11,48 | 10,94 | 8,65  | 7,61  | 9,62  | 11,08 |
| 78  | 12,30 | 10,89 | 10,36 | 8,07  | 7,03  | 8,86  | 10,4  |
| 79  | 11,69 | 10,32 | 9,80  | 7,52  | 6,48  | 8,12  | 9,73  |
| 80  | 11,10 | 9,77  | 9,25  | 6,99  | 5,96  | 7,42  | 9,1   |
| 81  | 10,52 | 9,22  | 8,72  | 6,48  | 5,47  | 6,75  | 8,48  |
| 82  | 9,96  | 8,70  | 8,21  | 6,00  | 5,00  | 6,11  | 7,89  |
| 83  | 9,41  | 8,19  | 7,72  | 5,54  | 4,56  | 5,50  | 7,33  |
| 84  | 8,88  | 7,70  | 7,24  | 5,11  | 4,15  | 4,94  | 6,79  |
| 85  | 8,36  | 7,22  | 6,78  | 4,70  | 3,77  | 4,41  | 6,28  |
| 86  | 7,86  | 6,76  | 6,33  | 4,31  | 3,42  | 3,92  | 5,8   |
| 87  | 7,37  | 6,32  | 5,90  | 3,94  | 3,09  | 3,47  | 5,36  |
| 88  | 6,90  | 5,89  | 5,49  | 3,60  | 2,78  | 3,05  | 4,94  |
| 89  | 6,44  | 5,48  | 5,10  | 3,28  | 2,50  | 2,67  | 4,55  |
| 90  | 6,00  | 5,08  | 4,72  | 2,98  | 2,24  | 2,33  | 4,2   |
| 91  | 5,57  | 4,70  | 4,36  | 2,70  | 2,00  | 2,03  | 3,87  |
| 92  | 5,16  | 4,34  | 4,01  | 2,43  | 1,78  | 1,76  | 3,58  |
| 93  | 4,77  | 3,99  | 3,68  | 2,19  | 1,59  | 1,51  | 3,3   |
| 94  | 4,39  | 3,66  | 3,37  | 1,97  | 1,40  | 1,30  | 3,04  |
| 95  | 4,03  | 3,35  | 3,07  | 1,76  | 1,24  | 1,11  | 2,81  |
| 96  | 3,68  | 3,05  | 2,79  | 1,57  | 1,09  | 0,95  | 2,6   |
| 97  | 3,36  | 2,76  | 2,53  | 1,39  | 0,96  | 0,81  | 2,41  |
| 98  | 3,04  | 2,49  | 2,28  | 1,23  | 0,84  | 0,68  | 2,25  |
| 99  | 2,74  | 2,24  | 2,04  | 1,08  | 0,73  | 0,57  | 2,09  |
| 100 | 2,46  | 2,00  | 1,82  | 0,95  | 0,63  | 0,48  | 1,96  |
| 101 | 2,20  | 1,78  | 1,61  | 0,82  | 0,54  | 0,40  | 1,84  |
| 102 | 1,95  | 1,57  | 1,42  | 0,71  | 0,46  | 0,33  | 1,74  |
| 103 | 1,71  | 1,37  | 1,24  | 0,61  | 0,39  | 0,28  | 1,64  |
| 104 | 1,49  | 1,19  | 1,07  | 0,52  | 0,33  | 0,23  | 1,56  |
| 105 | 1,29  | 1,03  | 0,92  | 0,44  | 0,28  | 0,18  | 1,49  |
| 106 | 1,10  | 0,87  | 0,78  | 0,37  | 0,23  | 0,15  | 1,43  |
| 107 | 0,93  | 0,73  | 0,66  | 0,30  | 0,19  | 0,12  | 1,37  |
| 108 | 0,77  | 0,61  | 0,54  | 0,24  | 0,15  | 0,09  | 1,33  |
| 109 | 0,63  | 0,49  | 0,44  | 0,19  | 0,12  | 0,07  | 1,29  |
| 110 | 0,50  | 0,39  | 0,35  | 0,15  | 0,09  | 0,05  | 1,27  |
| 111 | 0,39  | 0,30  | 0,27  | 0,11  | 0,07  | 0,04  |       |
| 112 | 0,29  | 0,22  | 0,20  | 0,08  | 0,05  | 0,03  |       |
| 113 | 0,21  | 0,16  | 0,14  | 0,06  | 0,03  | 0,02  |       |
| 114 | 0,14  | 0,10  | 0,09  | 0,04  | 0,02  | 0,01  |       |
| 115 | 0,08  | 0,06  | 0,05  | 0,02  | 0,01  | 0,01  |       |
| 116 | 0,04  | 0,03  | 0,03  | 0,01  | 0,01  | 0,00  |       |
| 117 | 0,01  | 0,01  | 0,01  | 0,00  | 0,00  | 0,00  |       |

Another indicator is the estimated maximum deterioration age (Max Det) which is stable and independent of the age level selected. The maximum deterioration age during recent years tends to coincide with the life expectancy at birth estimated with the classical techniques of constructing life tables. Both the DTR and the Max Det are quite good measures of the life expectancy now and in the future as is presented in Figure 8 where the mean values for the 15 countries studied are given.

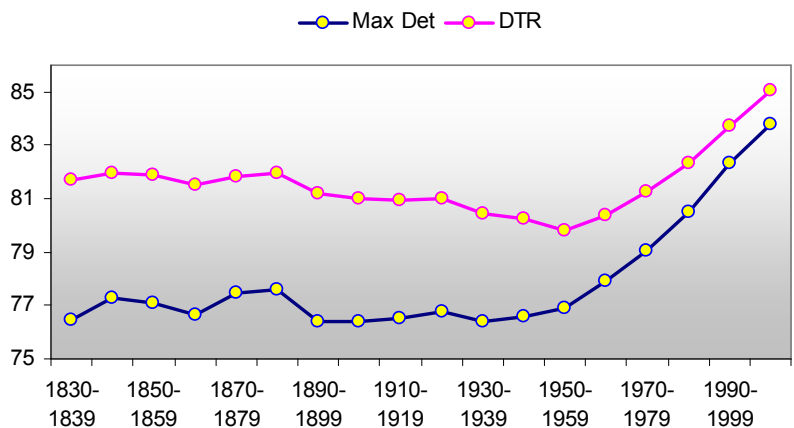

Fig. 8. Maximum deterioration age and DTR for 15 Countries

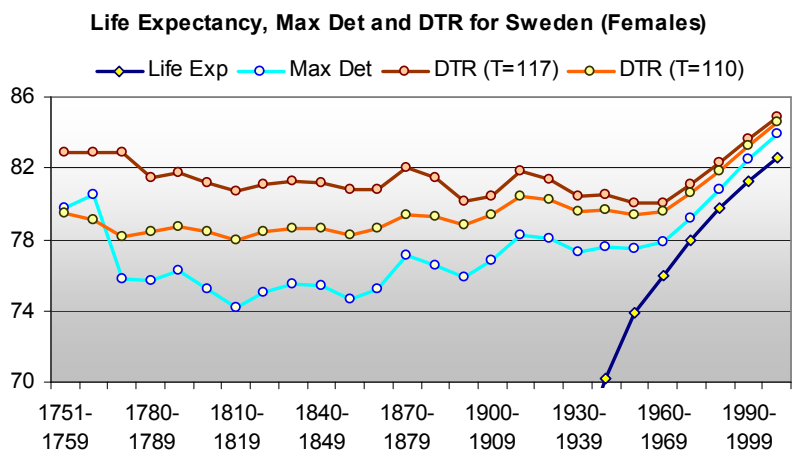

Figure 9. Life expectancy, Maximum deterioration age and DTR for Sweden, females

The influence of the life level $T$ for the estimation of life expectancy by the DTR system in Sweden (females) is illustrated in Figure 9. Two scenarios are selected for the estimation of the future life expectancy at birth. In the first a $T$=117 year level is accepted (dark brawn line) and in the second $T$=110. As it was expected the higher level for $T$ suggests a higher level for the life expectancy at birth via the DTR method. However, both scenarios tend to coincide in recent years something that it is quite useful in estimating the future trends for the life expectancy development (see Table VI). That it is important with the DTR system is that we can construct life tables for future dates thus doing forecasts. Instead with the Max Det we can have only an estimate for the future levels of life expectancy at birth but not for the life expectancy in other ages. The standard life expectancy at birth is presented (blue line) and the Max Deterioration points are also presented (light bleu line).

**TABLE VI**

| Year | Max Det | DTR |
|------|---------|------|
| 1830-1839 | 76,47 | 81,67 |
| 1840-1849 | 77,26 | 81,98 |
| 1850-1859 | 77,10 | 81,89 |
| 1860-1869 | 76,65 | 81,51 |
| 1870-1879 | 77,49 | 81,83 |
| 1880-1889 | 77,59 | 81,96 |
| 1890-1899 | 76,38 | 81,17 |
| 1900-1909 | 76,39 | 81,03 |
| 1910-1919 | 76,52 | 80,93 |
| 1920-1929 | 76,76 | 80,98 |
| 1930-1939 | 76,36 | 80,42 |
| 1940-1949 | 76,57 | 80,26 |
| 1950-1959 | 76,87 | 79,80 |
| 1960-1969 | 77,90 | 80,39 |
| 1970-1979 | 79,02 | 81,24 |
| 1980-1989 | 80,50 | 82,31 |
| 1990-1999 | 82,32 | 83,73 |
| 2000-2007 | 83,81 | 85,03 |

## The Halley Life Table

Edmund Halley published his famous paper in 1693. It was a pioneering study indicating of how a scientist of a high calibre could cope to a precisely selected data sets. Halley realized that to construct a life table from only mortality data was fusible only on the basis of a stationary population (Keyfitz and Caswell, 1977) by means of a population where births and deaths are almost equal and the incoming and outgoing people are limited. This was the case of the Breslau city in Silesia (now Wroclaw). The birth and death data sent to Halley gave him the opportunity to construct a life table and present his results in the paper on "An Estimate of the Degrees of the Mortality of Mankind, drawn from curious Tables of the Births and Funerals at the City of Breslau; with an Attempt to ascertain the Price of Annuities upon Lives". The same time period was also proposed a method for handling life tables by Graunt (1662). For more information on the history and the development of actuarial science see the related history by Haberman and Sibbett (1995).

The purpose of this chapter is first to use the Halley's life table data in order to construct a mortality curve by applying a stochastic model resulting from the first exit time theory. After applying the model and constructing the mortality curve we find the deterioration function for the specific population of Breslau at the years studied by Halley and thus making possible to find the maximum deterioration age and constructing a graph for the "vitality" of the population, a term proposed by Halley and used many years later by Strehler and Mildvan (1960) who also suggest the term "vitality" of a person in a stochastic modeling of the human life.

In Table VII the first two columns include the Breslau life table data from the Halley paper whereas in the third column we have constructed the deaths per age as the difference between two consecutive rows of the second column. The data from the third column are inserted into our Excel program of the fist exit time distribution function and the results are presented in Figure 10.

### TABLE VII

| Age | Persons (Total) | Persons | Age | Persons (Total) | Persons |
|-----|-----------------|---------|-----|-----------------|---------|
| 1 | 1000 | 145 | 51 | 335 | 11 |
| 2 | 855 | 57 | 52 | 324 | 11 |
| 3 | 798 | 38 | 53 | 313 | 11 |
| 4 | 760 | 28 | 54 | 302 | 10 |
| 5 | 732 | 22 | 55 | 292 | 10 |
| 6 | 710 | 18 | 56 | 282 | 10 |
| 7 | 692 | 12 | 57 | 272 | 10 |
| 8 | 680 | 10 | 58 | 262 | 10 |
| 9 | 670 | 9 | 59 | 252 | 10 |
| 10 | 661 | 8 | 60 | 242 | 10 |
| 11 | 653 | 7 | 61 | 232 | 10 |
| 12 | 646 | 6 | 62 | 222 | 10 |
| 13 | 640 | 6 | 63 | 212 | 10 |
| 14 | 634 | 6 | 64 | 202 | 10 |
| 15 | 628 | 6 | 65 | 192 | 10 |
| 16 | 622 | 6 | 66 | 182 | 10 |
| 17 | 616 | 6 | 67 | 172 | 10 |
| 18 | 610 | 6 | 68 | 162 | 10 |
| 19 | 604 | 6 | 69 | 152 | 10 |
| 20 | 598 | 6 | 70 | 142 | 11 |
| 21 | 592 | 6 | 71 | 131 | 11 |
| 22 | 586 | 7 | 72 | 120 | 11 |
| 23 | 579 | 6 | 73 | 109 | 11 |
| 24 | 573 | 6 | 74 | 98 | 10 |
| 25 | 567 | 7 | 75 | 88 | 10 |
| 26 | 560 | 7 | 76 | 78 | 10 |
| 27 | 553 | 7 | 77 | 68 | 10 |
| 28 | 546 | 7 | 78 | 58 | 9 |
| 29 | 539 | 8 | 79 | 49 | 8 |
| 30 | 531 | 8 | 80 | 41 | 7 |
| 31 | 523 | 8 | 81 | 34 | 6 |
| 32 | 515 | 8 | 82 | 28 | 5 |
| 33 | 507 | 8 | 83 | 23 | 3 |
| 34 | 499 | 9 | 84 | 20 | |
| 35 | 490 | 9 | 85 | | |
| 36 | 481 | 9 | 86 | | |
| 37 | 472 | 9 | 87 | | |
| 38 | 463 | 9 | 88 | | |
| 39 | 454 | 9 | 89 | | |
| 40 | 445 | 9 | 90 | | |
| 41 | 436 | 9 | 91 | | |
| 42 | 427 | 10 | 92 | | |
| 43 | 417 | 10 | 93 | | |
| 44 | 407 | 10 | 94 | | |
| 45 | 397 | 10 | 95 | | |
| 46 | 387 | 10 | 96 | | |
| 47 | 377 | 10 | 97 | | |
| 48 | 367 | 10 | 98 | | |
| 49 | 357 | 11 | 99 | | |
| 50 | 346 | 11 | 100 | | |

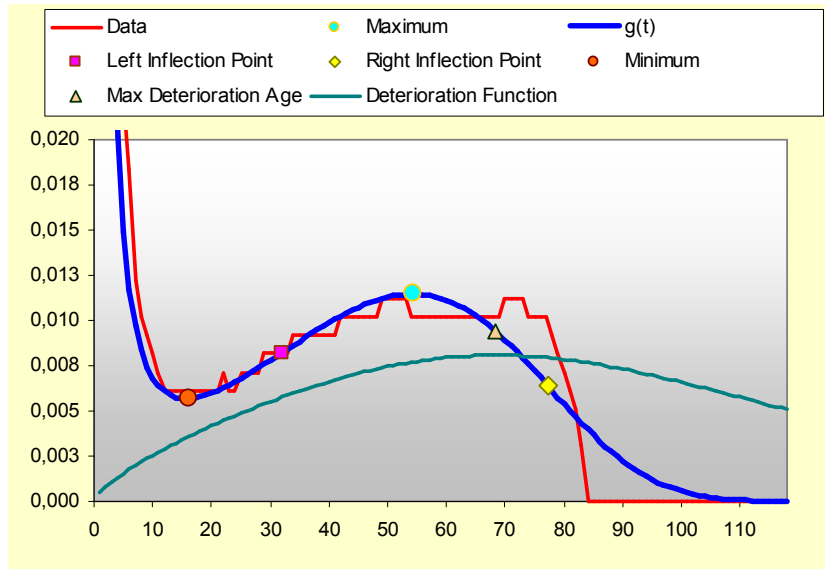

Fig. 10. Fit curve, data plot and deterioration curve for Halley data

The estimated best fit is presented with a blue line. The parameter estimates and the values for the characteristic points are given in the next Table VIII:

**TABLE VIII**

| Characteristic Points of Graph | Year | g(t) | g'(t) | Parameter |
|---|---|---|---|---|
| Maximum | 53,3 | 0,011457 | 0 | c = 2,72 |
| Left Inflection Point | 30,9 | 0,008182 | 0,0001081 | b = 0,03425 |
| Right Inflection Point | 76,2 | 0,006412 | -0,000470 | l = 0,25907 |
| Minimum | 14,6 | 0,005679 | 0 | k = 0,61215 |
| Maximum Deterioration Age | 67,4 | | | |

From the previous Figure and Table the estimated maximum death rate is at the age of 53,3 years, the right inflection point is at 76,2 years, the left inflection point is at 30,9 years and the minimum at 14,6 years. A very important characteristic of the health state of the population is given by estimating the age where the maximum deterioration takes place. This is estimated at the age of 67,39 years and it is the maximum of the deterioration function presented by a green curve in the graph.

## TABLE IX

DTR System (Halley, Breslaw Data)

| 1751-1759 Data | Data Normalised | IM Model Estimates | Year x | K(x) | K(x) Normalised | xK(x) | xK(x) Normalised | Survival curve | Life Expectancy |
|---|---|---|---|---|---|---|---|---|---|
| 145 | 0,1480 | 0,1535 | 0 | 0,0004834 | 0,00067177 | 0 | 0 | 1 | 79,07 |
| 57 | 0,0582 | 0,0554 | 1 | 0,0007963 | 0,00110653 | 0,0007963 | 1,705E-05 | 1 | 78,07 |
| 38 | 0,0388 | 0,0306 | 2 | 0,0010662 | 0,00148166 | 0,0021325 | 4,566E-05 | 0,99998295 | 77,07 |
| 28 | 0,0286 | 0,0202 | 3 | 0,0013116 | 0,00182264 | 0,0039348 | 8,4251E-05 | 0,99993729 | 76,07 |
| 22 | 0,0224 | 0,0149 | 4 | 0,0015402 | 0,00214027 | 0,0061607 | 0,00013191 | 0,99985304 | 75,07 |
| 18 | 0,0184 | 0,0117 | 5 | 0,0017562 | 0,00244044 | 0,0087809 | 0,00018802 | 0,99972113 | 74,07 |
| 12 | 0,0122 | 0,0097 | 6 | 0,0019623 | 0,0027268 | 0,0117735 | 0,00025209 | 0,99953311 | 73,07 |
| 10 | 0,0102 | 0,0084 | 7 | 0,0021602 | 0,00300181 | 0,0151211 | 0,00032377 | 0,99928102 | 72,07 |
| 9 | 0,0092 | 0,0074 | 8 | 0,0023512 | 0,00326723 | 0,0188093 | 0,00040274 | 0,99895725 | 71,07 |
| 8 | 0,0082 | 0,0068 | 9 | 0,0025362 | 0,00352437 | 0,0228259 | 0,00048874 | 0,99855451 | 70,07 |
| 7 | 0,0071 | 0,0064 | 10 | 0,002716 | 0,00377424 | 0,0271602 | 0,00058155 | 0,99806577 | 69,07 |
| 6 | 0,0061 | 0,0061 | 11 | 0,0028912 | 0,00401762 | 0,0318027 | 0,00068095 | 0,99748422 | 68,08 |
| 6 | 0,0061 | 0,0059 | 12 | 0,0030621 | 0,00425514 | 0,0367449 | 0,00078677 | 0,99680327 | 67,08 |
| 6 | 0,0061 | 0,0057 | 13 | 0,0032291 | 0,0044873 | 0,0419789 | 0,00089884 | 0,99601649 | 66,08 |
| 6 | 0,0061 | 0,0057 | 14 | 0,0033927 | 0,00471452 | 0,0474973 | 0,001017 | 0,99511765 | 65,09 |
| 6 | 0,0061 | 0,0057 | 15 | 0,0035529 | 0,00493715 | 0,0532931 | 0,0011411 | 0,99410065 | 64,09 |
| 6 | 0,0061 | 0,0057 | 16 | 0,00371 | 0,00515547 | 0,0593596 | 0,00127099 | 0,99295955 | 63,10 |
| 6 | 0,0061 | 0,0058 | 17 | 0,0038641 | 0,0053697 | 0,0656904 | 0,00140655 | 0,99168856 | 62,10 |
| 6 | 0,0061 | 0,0059 | 18 | 0,0040155 | 0,00558005 | 0,0722792 | 0,00154763 | 0,99028201 | 61,11 |
| 6 | 0,0061 | 0,0060 | 19 | 0,0041642 | 0,00578667 | 0,0791198 | 0,00169409 | 0,98873439 | 60,12 |
| 6 | 0,0061 | 0,0061 | 20 | 0,0043103 | 0,0059897 | 0,086206 | 0,00184582 | 0,98704029 | 59,13 |
| 7 | 0,0071 | 0,0063 | 21 | 0,0044539 | 0,00618923 | 0,0935316 | 0,00200268 | 0,98519447 | 58,15 |
| 6 | 0,0061 | 0,0064 | 22 | 0,004595 | 0,00638535 | 0,1010905 | 0,00216452 | 0,98319179 | 57,16 |
| 6 | 0,0061 | 0,0066 | 23 | 0,0047337 | 0,00657812 | 0,1088761 | 0,00233123 | 0,98102727 | 56,18 |
| 7 | 0,0071 | 0,0068 | 24 | 0,0048701 | 0,0067676 | 0,1168823 | 0,00250266 | 0,97869604 | 55,20 |
| 7 | 0,0071 | 0,0070 | 25 | 0,0050041 | 0,00695381 | 0,1251023 | 0,00267866 | 0,97619338 | 54,22 |
| 7 | 0,0071 | 0,0072 | 26 | 0,0051357 | 0,00713676 | 0,1335295 | 0,0028591 | 0,97351472 | 53,24 |
| 7 | 0,0071 | 0,0074 | 27 | 0,0052651 | 0,00731647 | 0,1421569 | 0,00304383 | 0,97065562 | 52,27 |
| 8 | 0,0082 | 0,0076 | 28 | 0,0053921 | 0,00749292 | 0,1509774 | 0,00323269 | 0,96761179 | 51,30 |
| 8 | 0,0082 | 0,0078 | 29 | 0,0055167 | 0,00766611 | 0,1599837 | 0,00342553 | 0,96437909 | 50,33 |
| 8 | 0,0082 | 0,0080 | 30 | 0,0056389 | 0,007836 | 0,1691681 | 0,00362219 | 0,96095356 | 49,37 |
| 8 | 0,0082 | 0,0082 | 31 | 0,0057588 | 0,00800257 | 0,1785229 | 0,00382249 | 0,95733137 | 48,40 |
| 8 | 0,0082 | 0,0084 | 32 | 0,0058762 | 0,00816577 | 0,1880398 | 0,00402626 | 0,95350888 | 47,45 |
| 9 | 0,0092 | 0,0086 | 33 | 0,0059912 | 0,00832555 | 0,1977106 | 0,00423333 | 0,94948262 | 46,49 |
| 9 | 0,0092 | 0,0088 | 34 | 0,0061037 | 0,00848187 | 0,2075264 | 0,00444351 | 0,94524929 | 45,54 |
| 9 | 0,0092 | 0,0091 | 35 | 0,0062137 | 0,00863466 | 0,2174784 | 0,0046566 | 0,94080578 | 44,60 |
| 9 | 0,0092 | 0,0093 | 36 | 0,006321 | 0,00878386 | 0,2275573 | 0,0048724 | 0,93614918 | 43,66 |
| 9 | 0,0092 | 0,0095 | 37 | 0,0064258 | 0,00892941 | 0,2377536 | 0,00509072 | 0,93127678 | 42,72 |
| 9 | 0,0092 | 0,0097 | 38 | 0,0065278 | 0,00907122 | 0,2480573 | 0,00531154 | 0,92618605 | 41,79 |
| 9 | 0,0092 | 0,0099 | 39 | 0,0066271 | 0,00920923 | 0,2584584 | 0,00553405 | 0,92087471 | 40,86 |
| 9 | 0,0092 | 0,0101 | 40 | 0,0067237 | 0,00934336 | 0,2689465 | 0,00575862 | 0,91534066 | 39,94 |
| 10 | 0,0102 | 0,0102 | 41 | 0,0068173 | 0,00947353 | 0,2795107 | 0,00598482 | 0,90958204 | 39,03 |
| 10 | 0,0102 | 0,0104 | 42 | 0,0069081 | 0,00959966 | 0,2901403 | 0,00621242 | 0,90359722 | 38,12 |
| 10 | 0,0102 | 0,0106 | 43 | 0,0069959 | 0,00972167 | 0,3008238 | 0,00644117 | 0,89738481 | 37,22 |
| 10 | 0,0102 | 0,0107 | 44 | 0,0070807 | 0,00983948 | 0,3115499 | 0,00667083 | 0,89094364 | 36,32 |
| 10 | 0,0102 | 0,0109 | 45 | 0,0071624 | 0,009953 | 0,3223068 | 0,00690116 | 0,88427281 | 35,43 |
| 10 | 0,0102 | 0,0110 | 46 | 0,0072409 | 0,01006216 | 0,3330826 | 0,00713189 | 0,87737165 | 34,54 |
| 10 | 0,0102 | 0,0111 | 47 | 0,0073163 | 0,01016688 | 0,3438651 | 0,00736276 | 0,87023976 | 33,67 |
| 11 | 0,0112 | 0,0112 | 48 | 0,0073884 | 0,01026706 | 0,3546421 | 0,00759351 | 0,862877 | 32,79 |
| 11 | 0,0112 | 0,0113 | 49 | 0,0074572 | 0,01036265 | 0,3654011 | 0,00782388 | 0,85528349 | 31,93 |
| 11 | 0,0112 | 0,0114 | 50 | 0,0075226 | 0,01045357 | 0,3761295 | 0,0080536 | 0,84745961 | 31,08 |
| 11 | 0,0112 | 0,0114 | 51 | 0,0075846 | 0,01053974 | 0,3868146 | 0,00828238 | 0,83940601 | 30,23 |
| 11 | 0,0112 | 0,0114 | 52 | 0,0076431 | 0,0106211 | 0,3974436 | 0,00850997 | 0,83112363 | 29,39 |
| 10 | 0,0102 | 0,0115 | 53 | 0,0076982 | 0,01069758 | 0,4080037 | 0,00873608 | 0,82261366 | 28,56 |
| 10 | 0,0102 | 0,0115 | 54 | 0,0077497 | 0,01076912 | 0,4184821 | 0,00896044 | 0,81387758 | 27,74 |
| 10 | 0,0102 | 0,0114 | 55 | 0,0077976 | 0,01083568 | 0,4288659 | 0,00918278 | 0,80491714 | 26,92 |
| 10 | 0,0102 | 0,0114 | 56 | 0,0078418 | 0,01089719 | 0,4391425 | 0,00940282 | 0,79573437 | 26,12 |
| 10 | 0,0102 | 0,0113 | 57 | 0,0078824 | 0,01095362 | 0,449299 | 0,00962028 | 0,78633155 | 25,32 |
| 10 | 0,0102 | 0,0112 | 58 | 0,0079194 | 0,01100493 | 0,459323 | 0,00983491 | 0,77671127 | 24,54 |
| 10 | 0,0102 | 0,0111 | 59 | 0,0079526 | 0,01105109 | 0,4692019 | 0,01004644 | 0,76687635 | 23,76 |
| 10 | 0,0102 | 0,0110 | 60 | 0,0079821 | 0,01109206 | 0,4789236 | 0,0102546 | 0,75682991 | 22,99 |
| 10 | 0,0102 | 0,0108 | 61 | 0,0080078 | 0,01112783 | 0,4884759 | 0,01045913 | 0,74657531 | 22,23 |
| 10 | 0,0102 | 0,0107 | 62 | 0,0080298 | 0,01115839 | 0,4978472 | 0,01065979 | 0,73611618 | 21,49 |
| 10 | 0,0102 | 0,0105 | 63 | 0,008048 | 0,01118373 | 0,5070259 | 0,01085632 | 0,7254564 | 20,75 |
| 10 | 0,0102 | 0,0103 | 64 | 0,0080625 | 0,01120386 | 0,5160009 | 0,01104849 | 0,71460008 | 20,03 |
| 10 | 0,0102 | 0,0100 | 65 | 0,0080733 | 0,01121878 | 0,5247614 | 0,01123607 | 0,70355158 | 19,31 |
| 10 | 0,0102 | 0,0098 | 66 | 0,0080803 | 0,01122851 | 0,5332968 | 0,01141883 | 0,69231552 | 18,61 |
| 10 | 0,0102 | 0,0095 | 67 | 0,0080835 | 0,01123308 | 0,5415973 | 0,01159655 | 0,68089669 | 17,92 |
| 10 | 0,0102 | 0,0092 | 68 | 0,0080831 | 0,01123251 | 0,5496531 | 0,01176904 | 0,66930014 | 17,24 |
| 11 | 0,0112 | 0,0089 | 69 | 0,0080791 | 0,01122686 | 0,5574552 | 0,0119361 | 0,6575311 | 16,57 |
| 11 | 0,0112 | 0,0086 | 70 | 0,0080714 | 0,01121615 | 0,5649951 | 0,01209754 | 0,645595 | 15,91 |
| 11 | 0,0112 | 0,0083 | 71 | 0,0080601 | 0,01120046 | 0,5722845 | 0,01225319 | 0,63349746 | 15,26 |
| 11 | 0,0112 | 0,0079 | 72 | 0,0080452 | 0,01117983 | 0,579256 | 0,01240289 | 0,62124426 | 14,63 |
| 10 | 0,0102 | 0,0076 | 73 | 0,0080269 | 0,01115435 | 0,5859626 | 0,01254649 | 0,60884137 | 14,01 |
| 10 | 0,0102 | 0,0072 | 74 | 0,0080051 | 0,01112408 | 0,5923778 | 0,01268385 | 0,59629488 | 13,40 |
| 10 | 0,0102 | 0,0069 | 75 | 0,0079799 | 0,01108912 | 0,5984959 | 0,01281485 | 0,58361102 | 12,80 |
| 10 | 0,0102 | 0,0065 | 76 | 0,0079515 | 0,01104955 | 0,6043116 | 0,01293938 | 0,57079617 | 12,22 |
| 9 | 0,0092 | 0,0061 | 77 | 0,0079197 | 0,01100547 | 0,6098205 | 0,01305733 | 0,55785679 | 11,65 |
| 8 | 0,0082 | 0,0057 | 78 | 0,0078849 | 0,01095698 | 0,6150184 | 0,01316863 | 0,54479946 | 11,09 |
| 7 | 0,0071 | 0,0054 | 79 | 0,0078469 | 0,01090418 | 0,619902 | 0,0132732 | 0,53163083 | 10,55 |
| 6 | 0,0061 | 0,0050 | 80 | 0,0078059 | 0,01084721 | 0,6244688 | 0,01337098 | 0,51835763 | 10,01 |
| 5 | 0,0051 | 0,0047 | 81 | 0,0077619 | 0,01078616 | 0,6287165 | 0,01346193 | 0,50498666 | 9,50 |
| 3 | 0,0031 | 0,0043 | 82 | 0,0077152 | 0,01072118 | 0,6326437 | 0,01354602 | 0,49152473 | 8,99 |
| 0 | 0,0000 | 0,0040 | 83 | 0,0076657 | 0,01065238 | 0,6362496 | 0,01362323 | 0,47797871 | 8,50 |
| 0 | 0,0000 | 0,0037 | 84 | 0,0076135 | 0,0105799 | 0,6395341 | 0,01369355 | 0,46435548 | 8,02 |
| 0 | 0,0000 | 0,0033 | 85 | 0,0075588 | 0,01050388 | 0,6424973 | 0,013757 | 0,45066193 | 7,56 |
| 0 | 0,0000 | 0,0030 | 86 | 0,0075016 | 0,01042445 | 0,6451404 | 0,01381359 | 0,43690493 | 7,11 |
| 0 | 0,0000 | 0,0028 | 87 | 0,0074421 | 0,01034175 | 0,6474649 | 0,01386337 | 0,42309133 | 6,67 |
| 0 | 0,0000 | 0,0025 | 88 | 0,0073804 | 0,01025594 | 0,6494729 | 0,01390636 | 0,40922797 | 6,25 |
| 0 | 0,0000 | 0,0022 | 89 | 0,0073165 | 0,01016716 | 0,651167 | 0,01394263 | 0,39532161 | 5,84 |
| 0 | 0,0000 | 0,0020 | 90 | 0,0072506 | 0,01007555 | 0,6525504 | 0,01397226 | 0,38137897 | 5,44 |
| 0 | 0,0000 | 0,0018 | 91 | 0,0071827 | 0,00998127 | 0,6536268 | 0,01399523 | 0,36740672 | 5,06 |
| 0 | 0,0000 | 0,0016 | 92 | 0,007113 | 0,00988446 | 0,6544002 | 0,01401186 | 0,35341142 | 4,69 |
| 0 | 0,0000 | 0,0014 | 93 | 0,0070417 | 0,00978527 | 0,6548754 | 0,01402204 | 0,33939955 | 4,34 |
| 0 | 0,0000 | 0,0012 | 94 | 0,0069687 | 0,00968386 | 0,6550573 | 0,01402593 | 0,32537751 | 4,00 |
| 0 | 0,0000 | 0,0010 | 95 | 0,0068942 | 0,00958038 | 0,6549513 | 0,01402366 | 0,31135158 | 3,68 |
| 0 | 0,0000 | 0,0009 | 96 | 0,0068184 | 0,00947496 | 0,6545632 | 0,01401535 | 0,29732792 | 3,36 |
| 0 | 0,0000 | 0,0008 | 97 | 0,0067412 | 0,00936777 | 0,6538991 | 0,01400113 | 0,28331257 | 3,07 |
| 0 | 0,0000 | 0,0007 | 98 | 0,0066629 | 0,00925894 | 0,6529656 | 0,01398114 | 0,26931143 | 2,78 |
| 0 | 0,0000 | 0,0006 | 99 | 0,0065835 | 0,00914863 | 0,6517693 | 0,01395553 | 0,25533029 | 2,51 |
| 0 | 0,0000 | 0,0005 | 100 | 0,0065032 | 0,00903696 | 0,6503172 | 0,01392444 | 0,24137476 | 2,26 |
| 0 | 0,0000 | 0,0004 | 101 | 0,0064219 | 0,00892409 | 0,6486166 | 0,01388802 | 0,22745032 | 2,02 |
| 0 | 0,0000 | 0,0003 | 102 | 0,0063399 | 0,00881014 | 0,6466749 | 0,01384645 | 0,2135623 | 1,79 |
| 0 | 0,0000 | 0,0003 | 103 | 0,0062573 | 0,00869526 | 0,6444997 | 0,01379988 | 0,19971585 | 1,58 |
| 0 | 0,0000 | 0,0002 | 104 | 0,006174 | 0,00857958 | 0,6420989 | 0,01374847 | 0,18591597 | 1,38 |
| 0 | 0,0000 | 0,0002 | 105 | 0,0060903 | 0,00846321 | 0,6394804 | 0,0136924 | 0,1721675 | 1,19 |
| 0 | 0,0000 | 0,0001 | 106 | 0,0060062 | 0,00834629 | 0,636652 | 0,01363184 | 0,1584751 | 1,02 |
| 0 | 0,0000 | 0,0001 | 107 | 0,0059217 | 0,00822894 | 0,6336221 | 0,01356697 | 0,14484325 | 0,86 |
| 0 | 0,0000 | 0,0001 | 108 | 0,005837 | 0,00811127 | 0,6303986 | 0,01349795 | 0,13127629 | 0,71 |
| 0 | 0,0000 | 0,0001 | 109 | 0,0057522 | 0,00799339 | 0,6269897 | 0,01342496 | 0,11777834 | 0,58 |
| 0 | 0,0000 | 0,0001 | 110 | 0,0056673 | 0,00787542 | 0,6234038 | 0,01334817 | 0,10435338 | 0,47 |
| 0 | 0,0000 | 0,0000 | 111 | 0,0055824 | 0,00775747 | 0,6196488 | 0,01326777 | 0,09100521 | 0,36 |
| 0 | 0,0000 | 0,0000 | 112 | 0,0054976 | 0,00763962 | 0,615733 | 0,01318393 | 0,07773744 | 0,27 |
| 0 | 0,0000 | 0,0000 | 113 | 0,005413 | 0,00752198 | 0,6116645 | 0,01309682 | 0,06455351 | 0,19 |
| 0 | 0,0000 | 0,0000 | 114 | 0,0053285 | 0,00740464 | 0,6074512 | 0,0130066 | 0,05145669 | 0,13 |
| 0 | 0,0000 | 0,0000 | 115 | 0,0052444 | 0,00728769 | 0,6031012 | 0,01291346 | 0,03845009 | 0,08 |
| 0 | 0,0000 | 0,0000 | 116 | 0,0051605 | 0,00717121 | 0,5986224 | 0,01281756 | 0,02553663 | 0,04 |
| 0 | 0,0000 | 0,0000 | 117 | 0,0050771 | 0,00705528 | 0,5940223 | 0,01271907 | 0,01271907 | 0,01 |
| | 1,0000 | | | 0,7196193 | 1 | 46,70 | 1 | 79,07 | |

The DTR system as presented earlier provides the last Life Table IX from which we can find a future life expectancy based on the deterioration function. The surprising result is that the estimated life expectancy is quite close to the values for Germany from 2000-2008 and higher than the related values for Poland (2005-2009) as provided by the Human Mortality Database. The estimates are presented in the next Figure 11 along with the estimates for the life expectancy estimated by the Breslau data (see the green line in the graph).

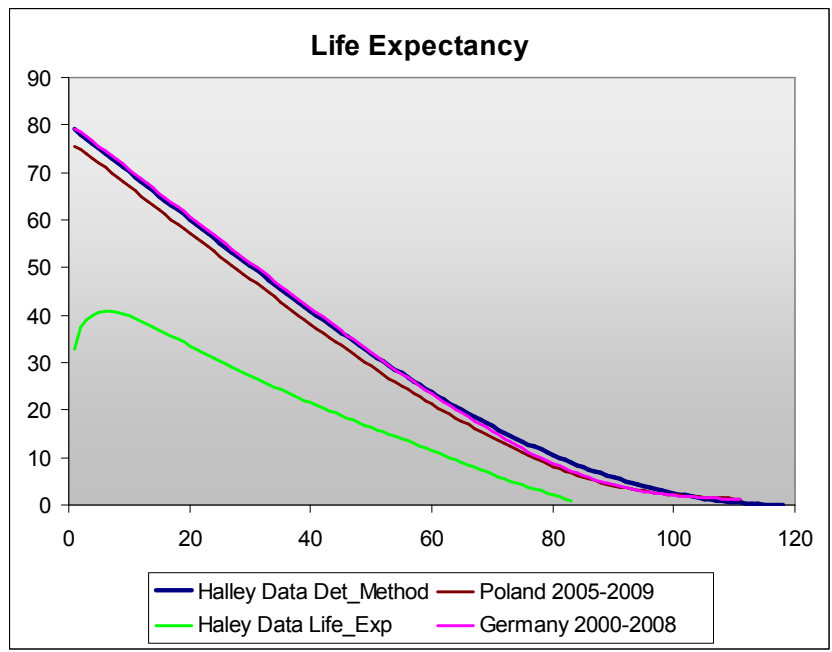

Fig. 11. Life expectancy curves for Breslau data

Illustration of the development of the maximum deterioration age from the Halley days until today is given in Figure 12. The maximum deterioration age for the Breslau data (1687-1691) was 67,39 years and continued to increase by 4,75 years per century.

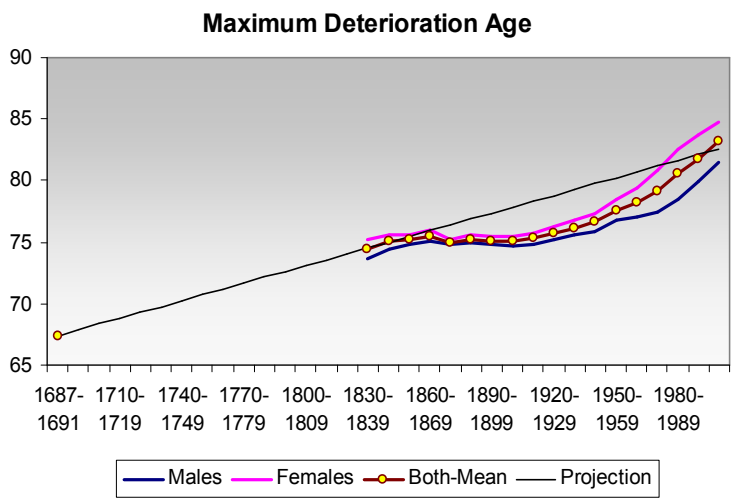

Fig. 12. Maximum Deterioration Age for various time periods

## The Program

A computer program (IM-model-DTR-Life_Tables) was developed to be able to make the necessary computations related to this paper. Furthermore the program estimates the life expectancy tables by based on the mortality and population data. The life expectancy is also estimated by based on the fitting curve thus making more accurate the related estimations. The program and the related theory can be found in the website: http://www.cmsim.net. The program is developed in Excel 2003 and it is very easy to use without any special tool.

## Conclusions

We have developed and applied a new theoretical framework for analyzing mortality data. The work starting several years ago was based on the stochastic theory and the derivation of a first exit time distribution function suitable for expressing the human mortality. Furthermore we have explored on how we could model the so-called "vitality" of a person or the opposite term the deterioration of an organism and to provide a function, the deterioration function, which could be useful for sociologists, police makers and the insurance people in making their estimates and plan the future.

## Acknowledgments

The data used can be downloaded from the Human Mortality Database at: http://www.mortality.org or from the statistical year-books of the countries studied.